\documentclass[final,5p,times,twocolumn]{elsarticle}

\usepackage{amssymb}
\usepackage{amsmath}    
\usepackage{graphicx}   
\usepackage{verbatim}   
\usepackage{color}      
\usepackage{hyperref}   
\usepackage{natbib}
\usepackage{float}
\usepackage{multirow}

\journal{Physics Letters B}

\begin{document}

\begin{frontmatter}

\title{Contribution of domain wall networks to the CMB power spectrum}

\author[DAM]{A. Lazanu\corref{cor1}}
\ead{A.Lazanu@damtp.cam.ac.uk}

\author[CM,CM2]{C. J. A. P. Martins}
\ead{Carlos.Martins@astro.up.pt}

\author[DAM]{E. P. S. Shellard}
\ead{E.P.S.Shellard@damtp.cam.ac.uk}

\cortext[cor1]{Corresponding author}
\address[DAM]{Centre for Theoretical Cosmology, Department of Applied Mathematics and Theoretical Physics, Wilberforce Road, Cambridge CB3 0WA, United Kingdom}
\address[CM]{Centro de Astrof\'{\i}sica, Universidade do Porto, Rua das Estrelas, 4150-762 Porto, Portugal}
\address[CM2]{Instituto de Astrof\'{\i}sica e Ci\^encias do Espa\c co, CAUP, Rua das Estrelas, 4150-762 Porto, Portugal}

\begin{abstract}
We use three domain wall simulations from the radiation era to the late time dark energy domination era based on the PRS algorithm to calculate the energy-momentum tensor components of domain wall networks in an expanding universe. Unequal time correlators in the radiation, matter and cosmological constant epochs are calculated using the scaling regime of each of the simulations. The CMB power spectrum of a network of domain walls is determined. The first ever quantitative constraint for the domain wall surface tension is obtained using a Markov chain Monte Carlo method; an energy scale of domain walls of 0.93 MeV, which is close but below the Zel'dovich bound, is determined.
\end{abstract}

\end{frontmatter}

\section{Introduction}
All inflationary theories involving Grand Unification Scale phase transitions in the early universe predict the existence of topological defects \cite{kibble76}. The inflationary field develops a spontaneous symmetry breaking phase, where various kinds of topological defects may form: textures, monopoles, strings, domain walls \cite{shbook}. Although some of these defects leave characteristic patterns in the Cosmic Microwave Background (CMB), no conclusive direct observation of defects has been yet found. Of these topological defects, cosmic strings have been the most studied in the literature, because they were once considered to be the primary sources of anisotropy of the CMB \cite{PhysRevLett.53.1016}. The contribution from cosmic strings and textures in the CMB has been recently investigated by the Planck probe \cite{planckstr}, where stringent constraints on these defects were found.

In the case of domain walls, the situation is different, because their energy density grows faster than that of radiation and matter as the universe is expanding and hence they would eventually become the dominant part of the energy of the universe \cite{1975JETP...40....1Z, Vilenkin1985263}. For this reason, their symmetry breaking scale is believed to be constrained to around 1 MeV. Nevertheless, their existence is predicted by various cosmological models which have discrete broken symmetries \cite{Linde1990353, PhysRevD.60.103504}. The purpose of this paper is to carry out a more quantitative derivation of this bound.

In this paper, we briefly introduce the equations governing the evolution of domain walls in the universe. We use the results of the simulations to calculate the energy-momentum tensor of domain walls for each time considered. We then determine the unequal time correlators (UETCs) of the stress-energy tensor components at high resolution and precision, relevant for the Planck satellite. We use two simulations, covering the entire period from recombination to late-time $\Lambda$ domination. We diagonalise the UETCs and use the eigenvectors as sources for the CMB fluctuations, thus obtaining the domain-wall angular power spectrum. We combine the results from the three epoch considered in order to obtain the final power spectrum. Finally, we use a Markov chain Monte-Carlo parameter estimation (COSMOMC) code with the latest CMB likelihoods in order to determine the first-ever constraints on the domain wall amplitudes.

\section{Domain walls equations of motion}

Domain walls are the simplest cosmological defects, as they can simply be described by a single scalar field $\phi$. One starts with the Lagrangian describing a discrete broken symmetry, that can be written as:
\begin{equation}
\mathcal{L}=-\frac{1}{4\pi}\left[\frac{1}{2}\phi_{,\alpha}\phi^{,\alpha}+V(\phi) \right]
\label{lagrangian-walls}
\end{equation}
where the field $\phi$ is real and the potential $V$ has at least two degenerate minima \cite{1989ApJ...347..590P}. The energy-momentum tensor of the walls network can then be expressed in terms of this Lagrangian as follows: 
\begin{equation}
T_{\mu\nu}=\frac{1}{4\pi}\left[\phi_{,\mu}\phi_{,\nu}-g_{\mu\nu} \left[\frac{1}{2}\phi_{,\alpha}\phi^{,\alpha}+V(\phi) \right] \right]
\end{equation}

In a Friedmann-Lema\^{i}tre-Robertson-Walker expanding universe the metric is taken to be:
\begin{equation}
g_{\mu\nu}=a^2\text{diag}(-1,1,1,1)
\label{metric}
\end{equation}
where the 0$^{th}$ dimension corresponds to conformal time. With respect to this metric, the components of the energy-momentum tensor become:
\begin{align}
T_{00}&=\frac{1}{4\pi}\left[\frac{1}{2}\phi'^2+\frac{1}{2} (\nabla \phi)^2 +a^2 V(\phi)\right] \\
\label{emt1}
T_{0i}&=\frac{1}{4\pi}\left[\phi'\partial_{i} \phi \right] \\
T_{ij}&=\frac{1}{4\pi}\left[\partial_i \phi \partial_j \phi + \delta_{ij} \left( \frac{1}{2}\phi'^2-\frac{1}{2} (\nabla \phi)^2 -a^2V(\phi) \right) \right]
\label{emt}
\end{align}
where prime denotes a derivative with respect to conformal time and the Laplacian is expressed in physical coordinates.
By applying the standard variational technique:
\begin{equation}
\frac{1}{\sqrt{-g}}\partial_\mu\left(\sqrt{-g}\frac{\partial \mathcal{L}}{\partial\left(\partial_\mu \phi\right)}\right)=\frac{\partial \mathcal{L}}{\partial \phi}
\end{equation}
for the Lagrange density (\ref{lagrangian-walls}) with metric (\ref{metric}) the equation of motion for $\phi$ is obtained:
\begin{equation}
\frac{\partial^2 \phi}{\partial \eta^2}+2\left(\frac{d \text{ln} a}{d \text{ln} \eta}\right)\frac{1}{\eta}\frac{\partial \phi}{\partial \eta}-\nabla^2 \phi=-a^2\frac{\partial V}{\partial \eta}
\label{eom-original}
\end{equation}
where $\eta$ is the conformal time. In the case of a power-law expansion of the universe, $a \propto t^\lambda$ and  $\frac{d \text{ln} a}{d \text{ln} \eta}=\frac{\lambda}{1-\lambda}$ has the value 1 in the radiation era and 2 in the matter era.

In order to calculate the stress-energy tensor components one has to first solve equation of motion (\ref{eom-original}) and then to substitute the solution into the corresponding equations (\ref{emt1})-(\ref{emt}). However in the comoving coordinates described above, the thickness of the walls decreases as $a^{-1}$ and, as Eq. (\ref{eom-original}) has to be solved numerically on a grid, the wall thickness quickly becomes smaller than the grid spacing. This problem can be overcome by modifying Eq. (\ref{eom-original}) \cite{1989ApJ...347..590P} to:
\begin{equation}
\frac{\partial^2 \phi}{\partial \eta^2}+\alpha\left(\frac{d \text{ln} a}{d \text{ln} \eta}\right)\frac{1}{\eta}\frac{\partial \phi}{\partial \eta}-\nabla^2 \phi=-a^\beta\frac{\partial V}{\partial \eta}
\label{eom-modificat}
\end{equation}
The unmodified equation of motion corresponds to $\alpha=\beta=2$. However, taking the coefficients to be $\alpha=3$ and $\beta=0$, one solves the problem of wall thinning, as the walls would have constant thickness in comoving coordinates (by modifying $\beta$) and would also maintain energy-momentum conservation (by modifying $\alpha$ as well). The procedure is called the Press-Ryden-Spergel (PRS) algorithm, after the names of the authors in \cite{1989ApJ...347..590P}. This equation can now be solved numerically on a grid using a finite difference scheme as follows:

\begin{align}
\label{finite-dif1}
\delta&\equiv\frac{1}{2}\alpha\frac{\Delta \eta}{\eta}\frac{d \text{ln}a}{d \text{ln}\eta} \\
\left(\nabla^2\phi\right)_{ijk}&=\phi_{i+1,j,k}+\phi_{i-1,j,k}+\phi_{i,j+1,k}+ \\
& +\phi_{i,j-1,k}+\phi_{i,j,k+1}+\phi_{i+1,j,k-1}-6\phi_{i,j,k} \nonumber \\ 
\dot{\phi}_{ijk}^{n+\frac{1}{2}}&=\frac{\left(1-\delta\right)\dot{\phi}_{ijk}^{n-\frac{1}{2}}+\Delta\eta\left(\nabla^2\phi_{ijk}^n-a^\beta\frac{\partial V}{\partial \phi_{ijk}^n}\right)}{1+\delta} \\
\phi_{ijk}^{n+1}&=\phi_{ijk}^{n}+\Delta\eta\dot{\phi}_{ijk}^{n+\frac{1}{2}}
\label{finite-dif}
\end{align}
where dots denote derivatives with respect to conformal time. The effects of this modification of the equations of motion have been studied in detail by various authors, most recently in \cite{PhysRevD.84.103523}, and no significant effect on the domain wall characteristics has been observed.

These equations use the assumption that the domain walls always have a small contribution on the overall energy density of the universe. This assumption is based on the fact that no direct signs of domain walls have been observed.  Hence their contribution to the matter perturbations can be treated as a first order approximation in perturbation theory. Therefore, at this order, their evolution does not significantly affect the expansion of the universe and hence we can safely use a power law expansion rate for radiation and matter epochs. 

\section{Formalism for calculating the power spectrum}
\label{formal}
After obtaining the field $\phi$ and its time derivative we use Eqs. (\ref{emt1})-(\ref{emt}) to calculate the energy-momentum tensor, and then we project it onto a 3-dimensional grid as in Ref. \cite{string2014}. This stress-energy tensor is decomposed into its scalar, vector and tensor parts, and as in the case of the strings, we chose to use two scalar components ($\Theta_{00}$, $\Theta^S$), one vector and one tensor components.  We have chosen these particular variables, as they appear in the matter perturbation equation in the synchronous gauge. These get modified by the presence of domain walls as follows:

\begin{eqnarray}
\label{modif2}
6kh^{- \prime}=4\pi G a^2 \sum_i \left( \rho_i+p_i\right)v_i-\frac{4\pi G}{k} \Theta^D\\
\ddot{h}^S+2\frac{a'}{a} \dot{h}^S-12k^2 h^-=16 \pi G \left(a^2 p \Sigma^S+\Theta^S \right) \\
\ddot{h}^V+2\frac{a'}{a} \dot{h}^V=16 \pi G \left(a^2 p \Sigma^V+\Theta^V \right) \\
\ddot{h}^T+2\frac{a'}{a} \dot{h}^T+k^2 h^T=16 \pi G \left(a^2 p \Sigma^T+\Theta^T \right)
\label{modif3}
\end{eqnarray}
where $h^-=h-h^S$ and $\Theta^D$ satisfies the equation
\begin{equation}
\dot{\Theta}^D=\Theta^D \left(-2\frac{\dot{a}}{a}-\frac{k^2}{3\frac{\dot{a}}{a}} \right)-\frac{k^2}{3} \left(2\Theta^S-\Theta_{00}-\frac{\dot{\Theta}_{00}}{\frac{\dot{a}}{a}} \right)
\label{modif}
\end{equation}

Domain walls are active sources, and compared to inflationary fluctuations which are only generated at the surface of last scattering, they continuously source the metric perturbations. This makes direct CMB calculations impossible due to the huge amount of data generated. Fortunately, one may use UETCs of domain walls which contain all the relevant information at first order in perturbation theory. These UETCs represent the 2-point correlation function of different components of the energy-momentum tensor. This quantity has been studied extensively and has been used in the computation of the power spectrum for cosmic strings \cite{string2014, turok-causality, contaldi}. For domain walls, in this particular gauge, we need three scalar UETCs ($\langle \Theta_{00}\Theta_{00}\rangle$, $\langle \Theta^S\Theta^S \rangle$, $\langle \Theta_{00}\Theta^S \rangle$), one vector and one tensor UETCs. In fact, we have taken two vector and two tensor components and checked that they have the same auto-correlators and that their cross-correlators vanish. This is due to statistical isotropy \cite{hindmarsh2006}.

As domain walls have a different scaling law compared to cosmic strings, the UETCs computed directly are non-scaling even in a purely radiation or matter epoch. Therefore, in order to be able to use them in a Boltzmann code, we have to fix the scaling. In each epoch, the scaling behaviour is achieved by considering the quantity:
\begin{equation}
C\left(k\tau_1,k\tau_2 \right)=\frac{1}{\sqrt{\tau_1 \tau_2}} \langle \Theta(\textbf{k},\tau_1) \Theta(\textbf{k},\tau_2)\rangle
\label{UETCv2}
\end{equation}
where $\Theta$ corresponds to a generic component of the energy-momentum tensor. For the cases considered, these are positive definite functions, and hence they can be expressed in terms of their eigenvectors and positive eigenvalues as follows:
\begin{align}
C\left(k\tau_1,k\tau_2 \right)&=\sum_i \lambda_i v_i(k\tau_1)^T v_i(k\tau_2)  \nonumber \\ 
&= \sum_i w_i(k\tau_1)^T w_i(k\tau_2)  
\end{align}

with $w_i=\sqrt{\lambda_i} v_i$
Then one would have to substitute the energy-momentum tensor component with
\begin{equation}
\Theta\left(\textbf{k},\tau \right) \to \sqrt{\tau} w_i\left(k\tau \right)=\sqrt{\tau}\sqrt{\lambda_i}v_i\left(k\tau \right)
\label{subst}
\end{equation}
due to the fact that we have divided by $\sqrt{\tau}$ in Eq. \ref{UETCv2}. For the vector and tensor cases this can be done straightforwardly, but for scalars we have to create a matrix formed by all the scalar components and then diagonalise it:
\begin{equation}
\begin{pmatrix}
\langle \Theta_{00}\Theta_{00} \rangle & \langle \Theta_{00}\Theta^S \rangle \\ 
\langle \Theta^S\Theta_{00} \rangle & \langle \Theta^S\Theta^S \rangle 
\end{pmatrix}
\end{equation}
The first half of each of the eigenvectors corresponds to the $\Theta_{00}$ part, while the second one to $\Theta^S$. The eigenvalues are common to both. The resulting eigenvectors are used as sources in an Einstein-Boltzmann eigensolver separately for scalars, vectors and tensors and the results are then summed up in order to obtain the total angular power spectrum components.

This procedure works for a simulation accross both the radiation and the matter era if one assumes that the same unequal time correlators are valid in both epochs. However this is not the case and hence we use the UETCs only in their time range of validity. The procedure is described in greater detail in Ref. \cite{string2014}. For example, when calculating the contribution from the matter era, instead of making the substitution from Eq. (\ref{subst}), we perform the following transformation:
\begin{equation}
\label{split}
\Theta\left(\textbf{k},\tau \right) \to  \left\{
\begin{array}{cl}
0 & \text{if } \tau \in \text{radiation era} \\
\sqrt{\tau}\sqrt{\lambda_i}v_i\left(k\tau \right) & \text{if } \tau \in \text{matter era}\\
\end{array} \right.
\end{equation}
and similarly in the case of the radiation era. We go from one regime to the other by using a suitable smoothing function to avoid discontinuities and numerical problems. The total power spectrum is the sum of the power spectra obtained from the two eras. Although the procedure is not exact, in \cite{string2014} we show that indeed the errors are small.
Because domain walls are active sources and the fact that they are uncorrelated with the primordial fluctuations, one can calculate their power spectrum separately from the inflationary one and sum up the results at the end. This makes the calculation of the domain wall power spectrum easier, as we don't have to worry about other cosmological sources. 

In the cosmological constant era it is not possible to use the exponential expansion directly, because the expansion rate does not have a simple form as in the radiation and matter eras. However, it is important to use a late-time simulation, because as the growth of the domain walls density is greater than the background, their power spectrum is going to be dominated by the late-time contribution. We propose an approximation to the consmological constant epoch by considering the universe to be expanding with an effective power law, and we determine the power $\lambda$ of an expansion rate $a \propto t^{\lambda}$ that has the same slope today, as tha actual expansion rate. We then generalise Eq. (\ref{split}) to include the last simulation as well. We use both the two- and three-era calculations to determine the constraints on the domain walls, and then show that the change due to the last epoch is small.

\section{Simulations}
For our domain wall numerical simulations we use a code based on the PRS algorithm \cite{1989ApJ...347..590P} with the diagnostic tools introduced in \cite{PhysRevD.71.083509}. This has been successively parallelised and optimised to exploit recent HPC developments; the more recent version of the code is described in \cite{PhysRevD.84.103523}, and a forthcoming publication will describe further developments.

In order to solve Eqs. (\ref{finite-dif1})-(\ref{finite-dif}), we assume the following numerical values of the parameters involved:
\begin{align}
\alpha&=3 \\
\beta&=0 \\
V(\phi)&=\frac{\pi^2}{50}\left(\phi^2-1\right)^2
\end{align}
This corresponds to an initial wall thickness of $W_0=10$.

\begin{figure}[!htb]
\begin{center}$
\begin{array}{c}
\includegraphics[width=2.6in]{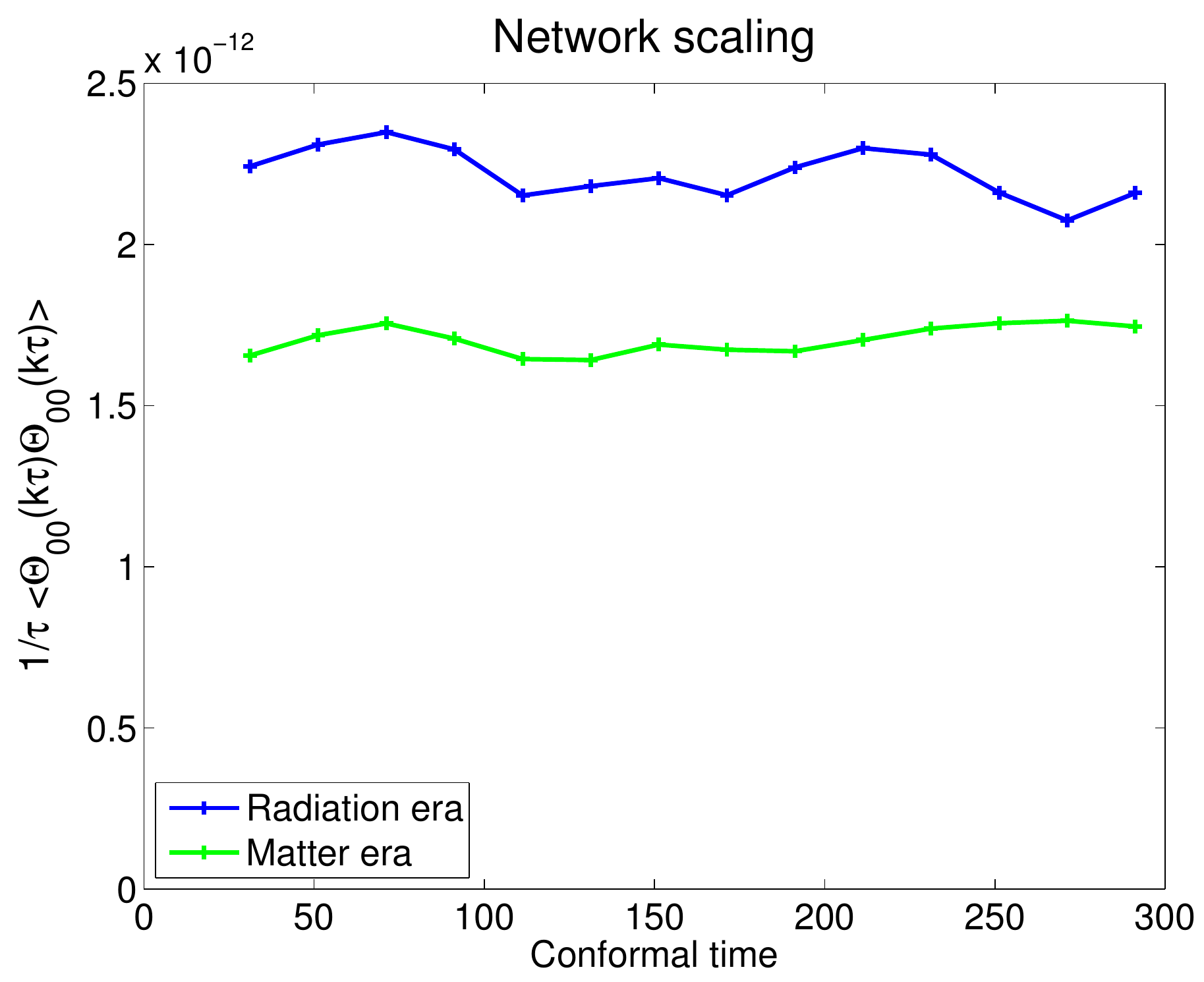} \\
\end{array}$
\caption{Scaling behaviour of the UETC in the radiation (blue) and matter (green) eras}
\label{scaling-fig}
\end{center}
\end{figure}

The grid spacing is taken to be $\Delta_x=1$ and the initial conditions are such that $\phi$ is taking random values on the grid between -1 and 1 and $\dot{\phi}=0$ everywhere.
We ran 3 three-dimensional simulations, with a box size of  $1024^3$ points, one in radiation era ($\lambda=1/2$), one in the matter era ($\lambda=2/3$) and one in in late-tiem $\Lambda$-dominated era.

The energy-momentum tensor of the domain wall network can be evaluated at any time, but we are only interested in the scaling regime of the simulation for the calculation of UETCs. The regime where the network exhibits such behaviour has been investigated in detail in \cite{PhysRevD.84.103523, Leite2013740}. In Figure \ref{scaling-fig} we show the peaks of the $\langle\Theta_{00}\Theta_{00}\rangle$ unequal time correlators calculated using Eq. (\ref{UETCv2}), with each of them centred around the value on the \textit{x}-axis. This shows how good the scaling is in this regime.

\begin{figure}[!htb]
\begin{center}$
\begin{array}{cc}
\includegraphics[width=1.5in]{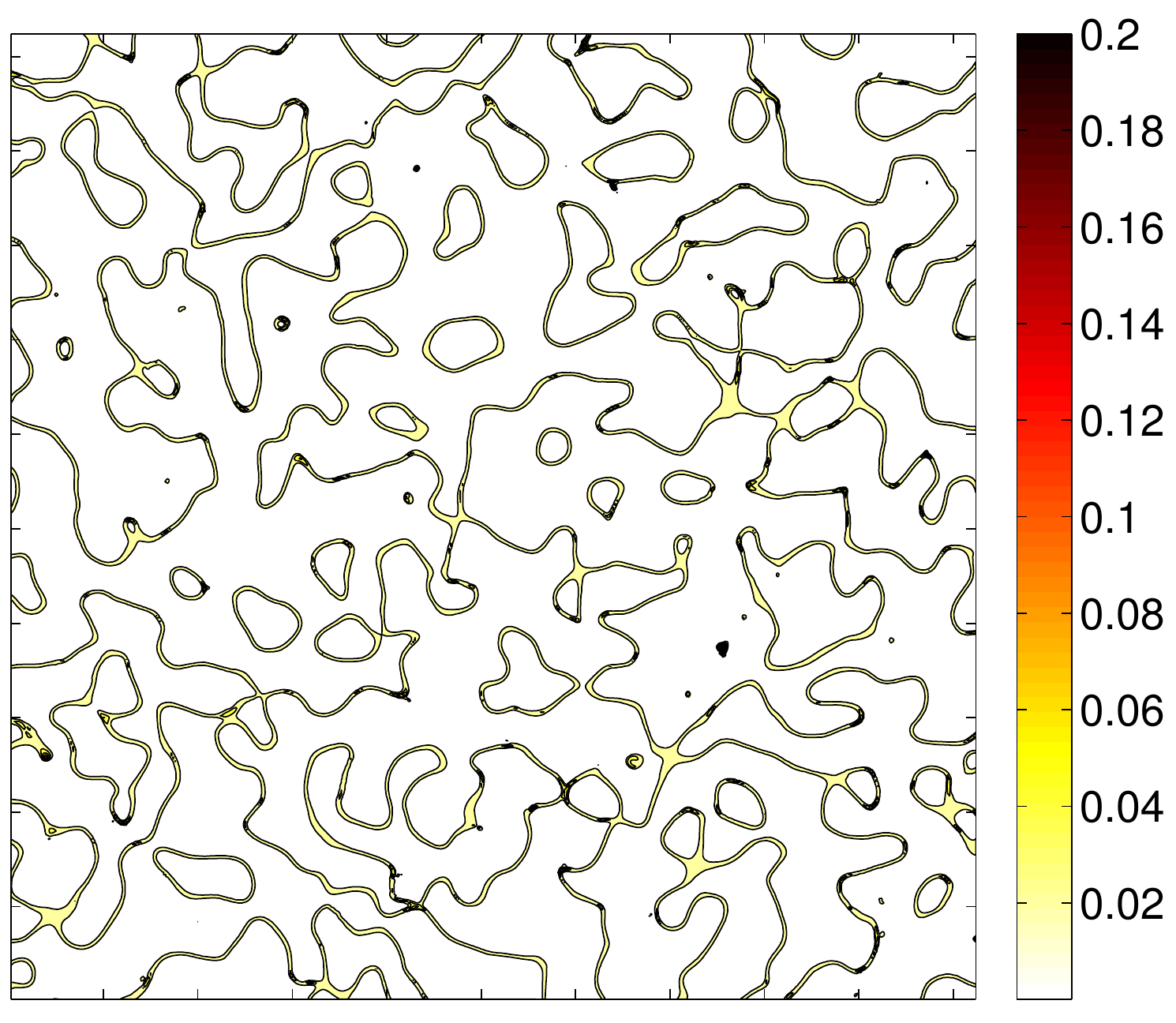} &
\includegraphics[width=1.5in]{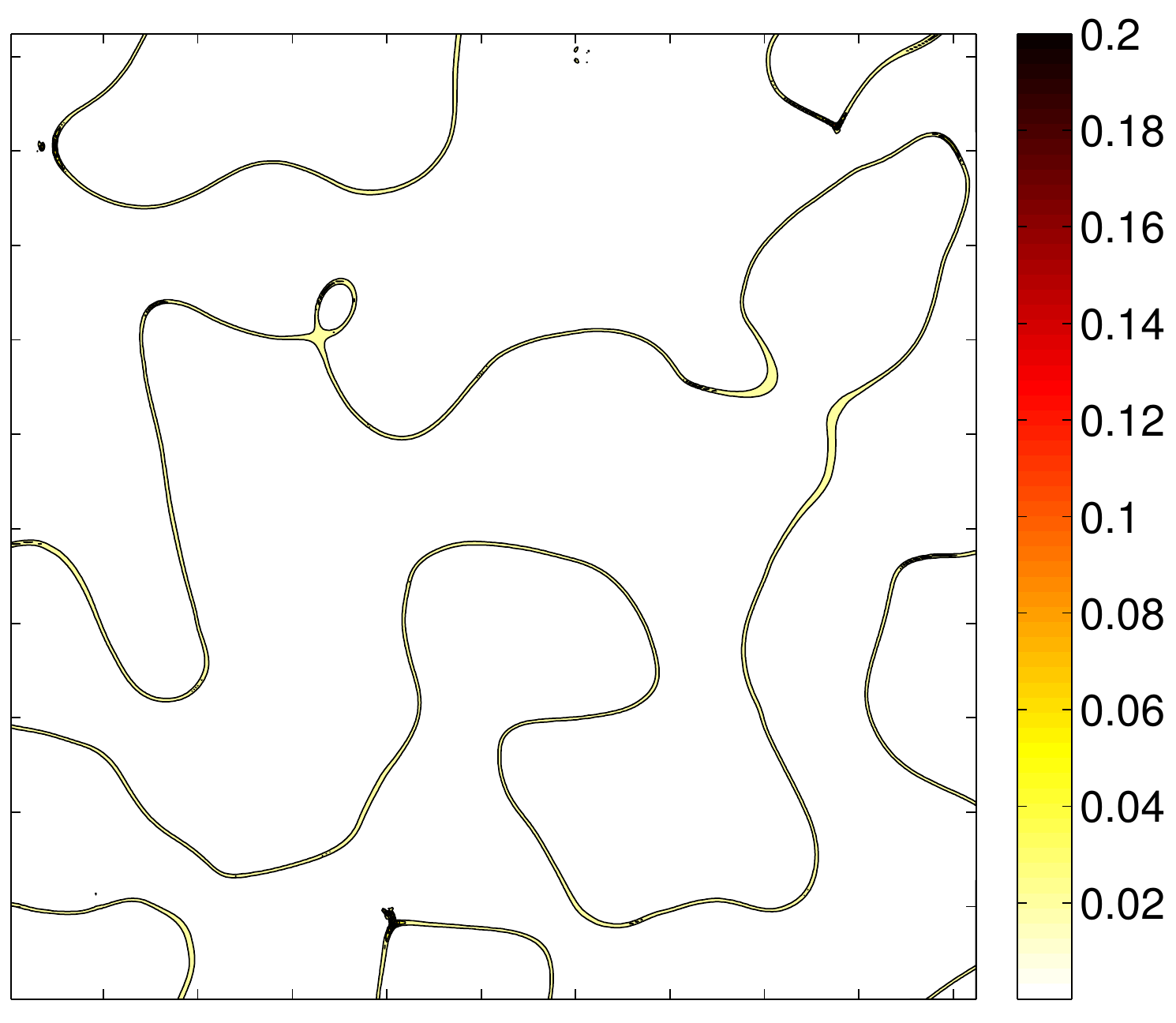} \\
\includegraphics[width=1.5in]{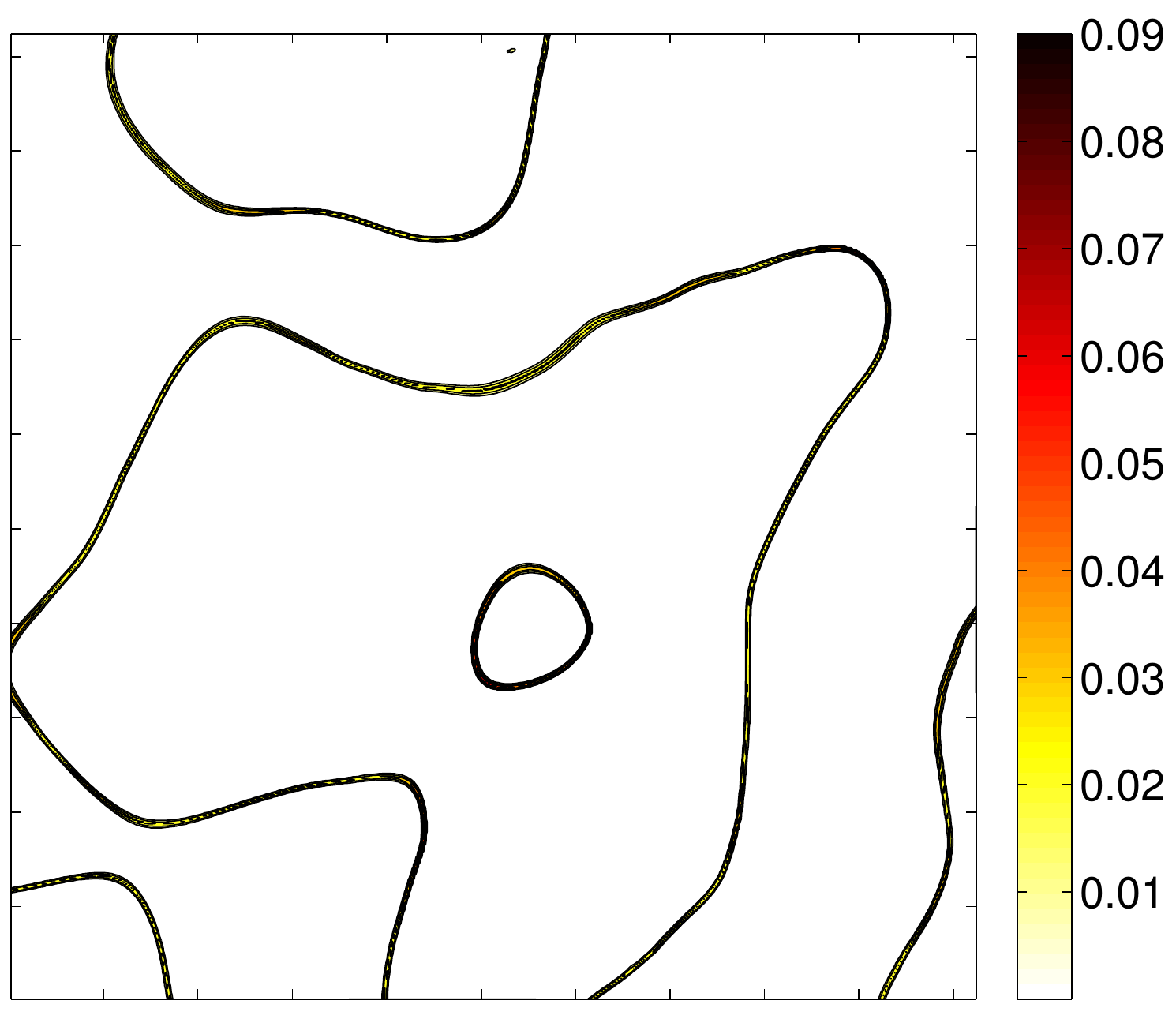} &
\includegraphics[width=1.5in]{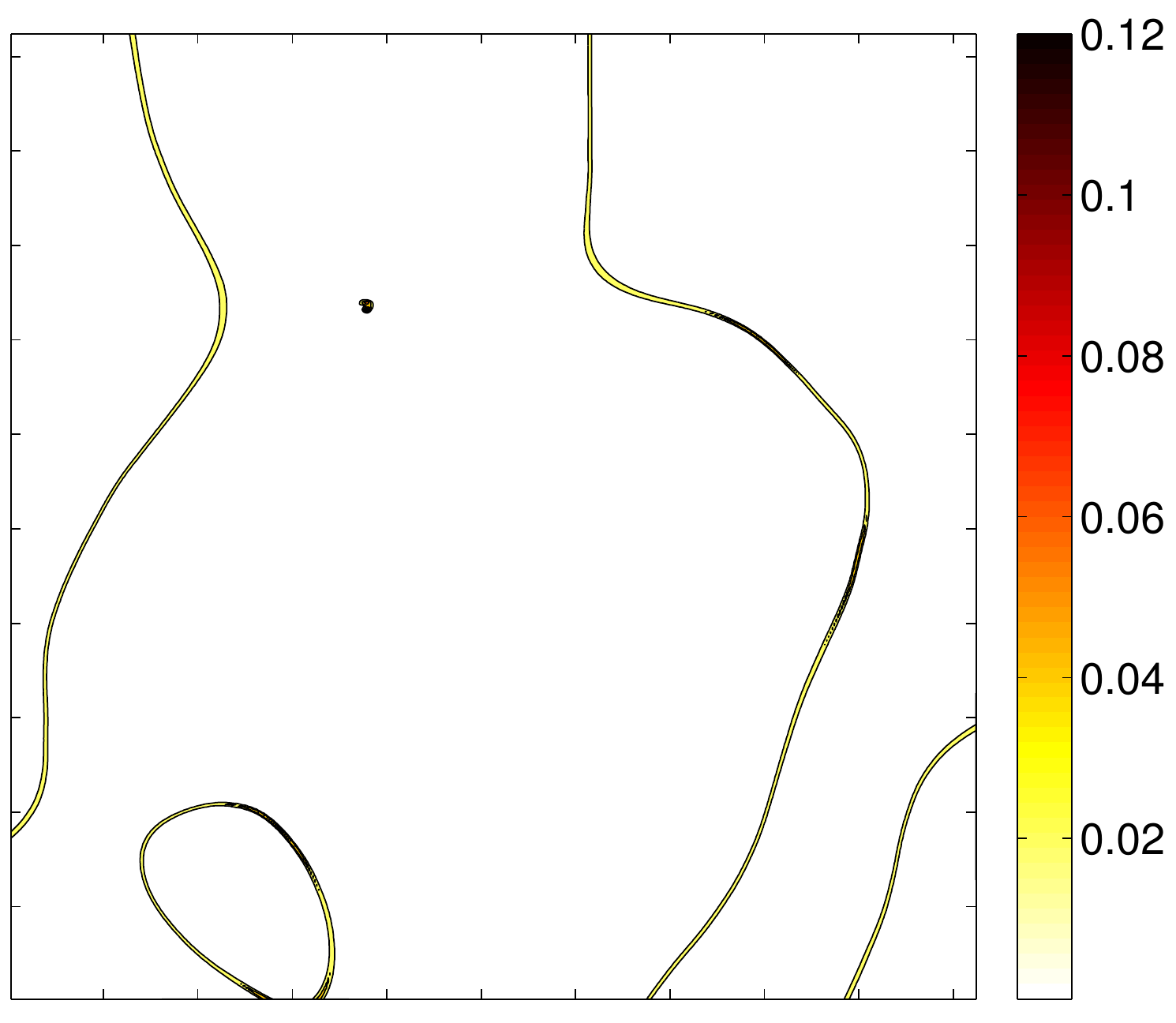} 
\end{array}$
\caption{Two-dimensional slices of the domain wall network during the scaling regime in the matter era (from left to right and top to bottom - roughly equal conformal time steps between the beginning and end of simulation)}
\label{2dslice}
\end{center}
\end{figure}

In Figure \ref{2dslice} we have plotted four 2-dimensional slices through the domain wall network at different times of the simulation, and these show how the network becomes less dense over time.

For the $\Lambda$-era, taking into account that the contribution today would be the most important, we consider a simulation with $\lambda=H_0 t_0$, where $t_0$ is the age of the universe. Using the values of the cosmological parameters from Ref. \cite{planckres}, we take an average value of $\lambda=0.95$ between various likelihoods.

In a universe with $a \propto t^\lambda$, the physical density of domain walls can be expressed as:
\begin{equation}
\rho=\left(1-\lambda \right) \left(\frac{A\eta}{V} \right) \frac{\sigma}{ct}
\label{physdens}
\end{equation}

Using the units in the simulation, we are obtaining the product of the first two terms in Eq. (\ref{physdens}). The domination of the power spectrum by the matter era contribution makes it possible to safely use $\lambda = \frac{2}{3}$ without introducing significant errors. In the matter era, for the scaling regime we took $\frac{A\eta}{V}=1.93$. Using Eq. (\ref{physdens}) and the fact that the background density of the universe is given by $\bar{\rho}=\frac{1}{6 \pi Gt^2}$. Hence, the coefficient multiplying the power spectrum today is given by:
\begin{equation}
\left(\frac{6 \pi G t_0 \sigma}{c}\right)^2
\end{equation}

\section{Results}

\begin{figure*}[!htb]
\begin{center}$
\begin{array}{cccc}
\includegraphics[width=1.63in]{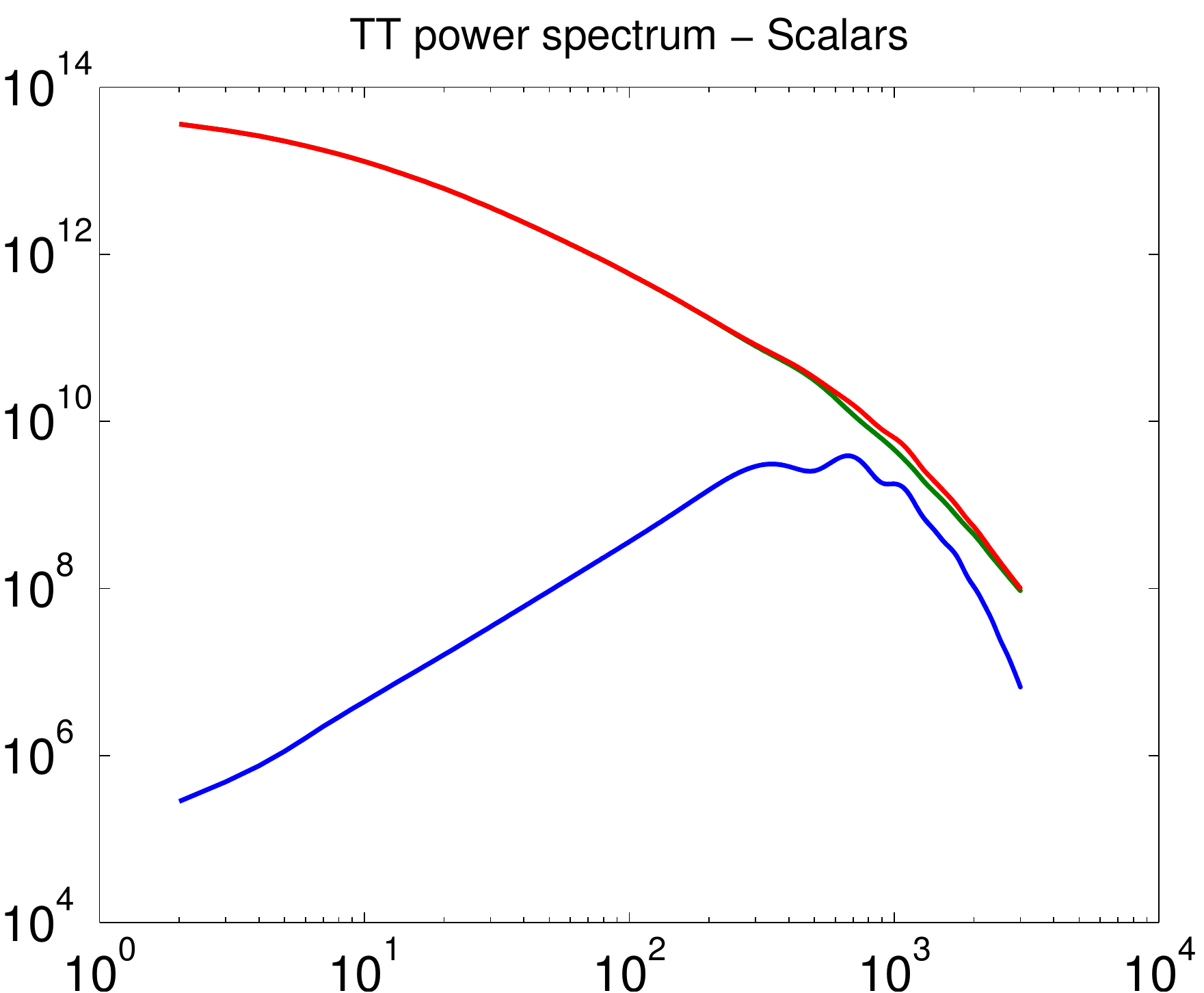} &
\includegraphics[width=1.63in]{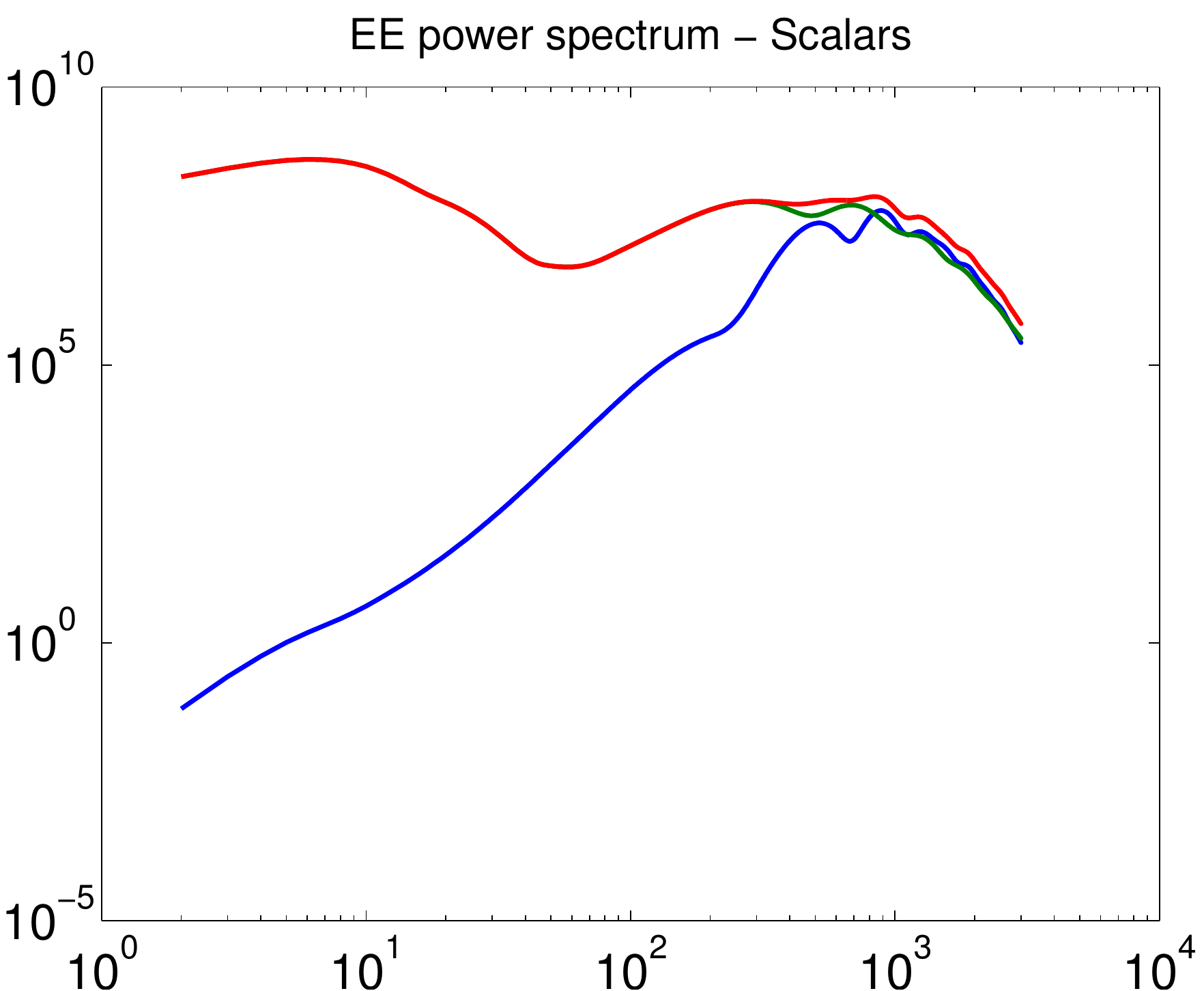} &
\includegraphics[width=1.63in]{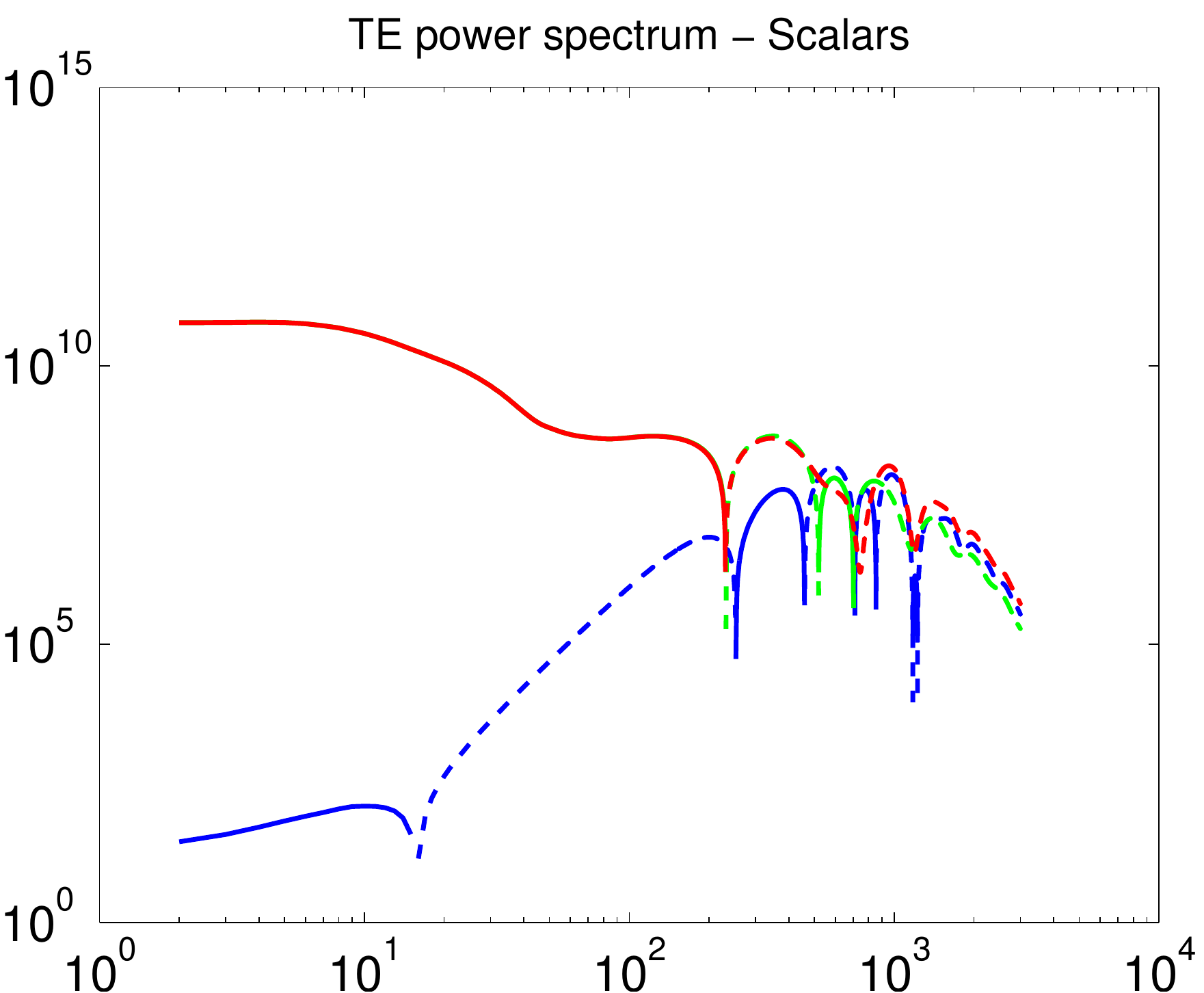} &
\includegraphics[width=1.63in]{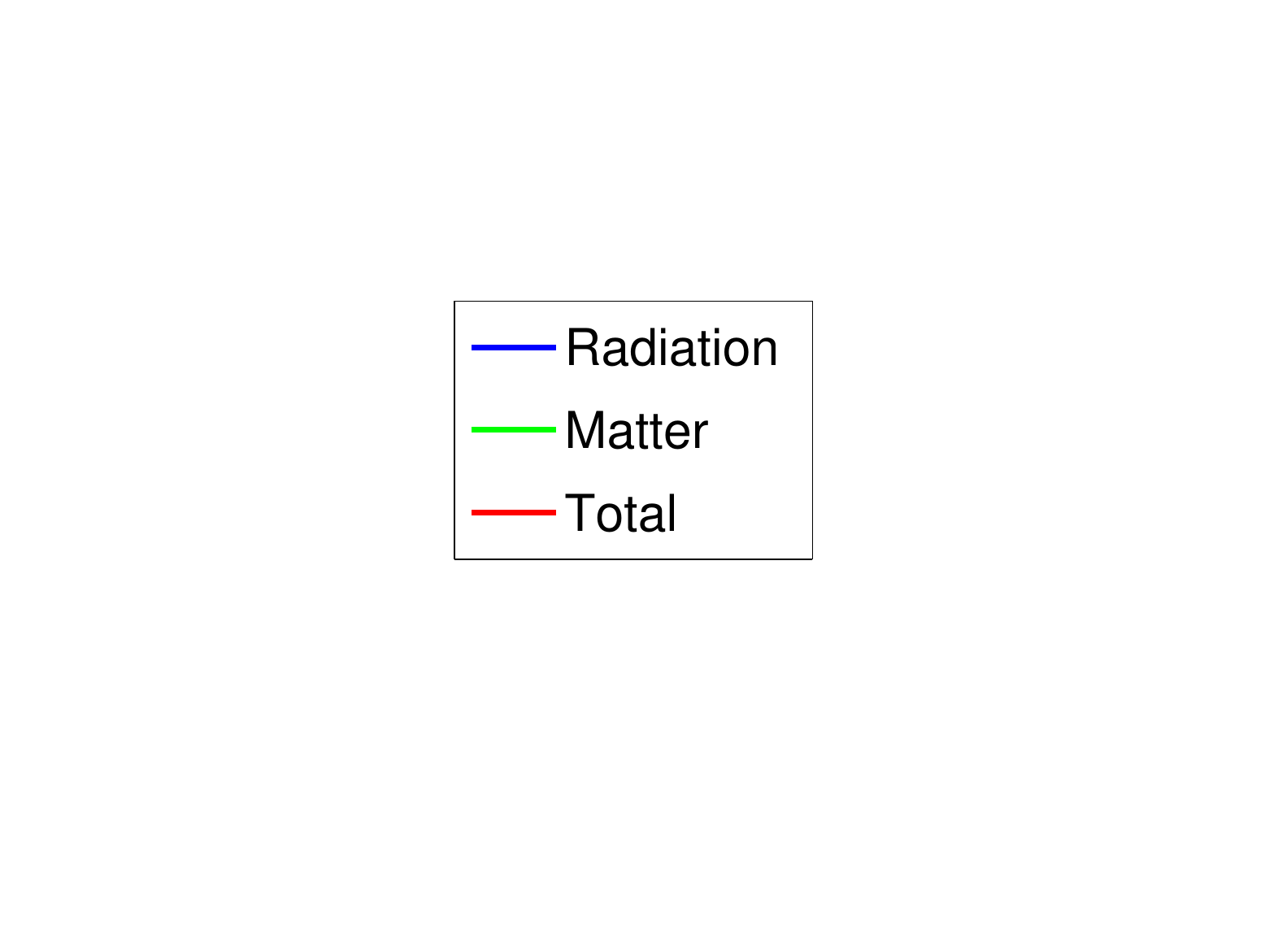} \\
\includegraphics[width=1.63in]{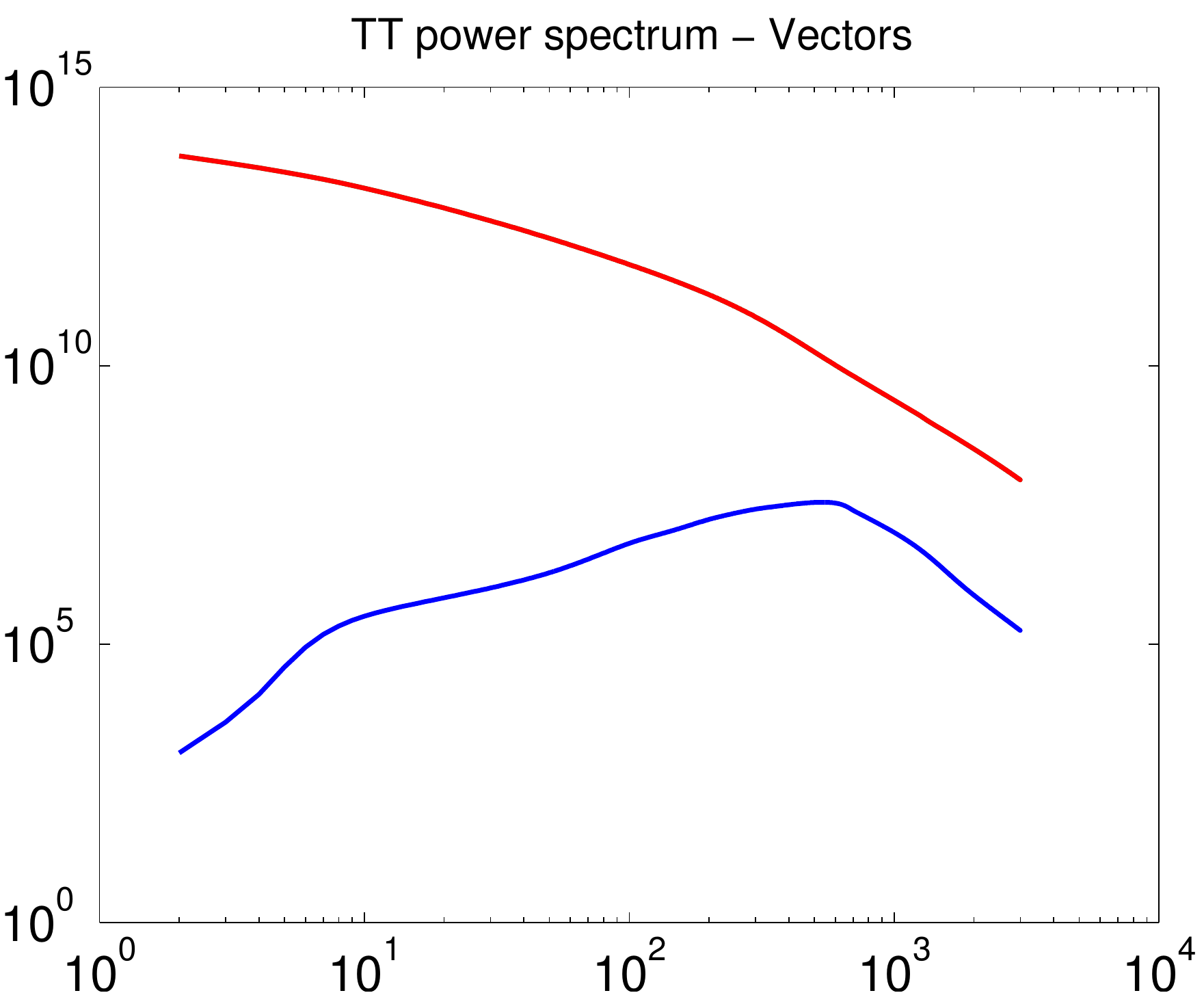} &
\includegraphics[width=1.63in]{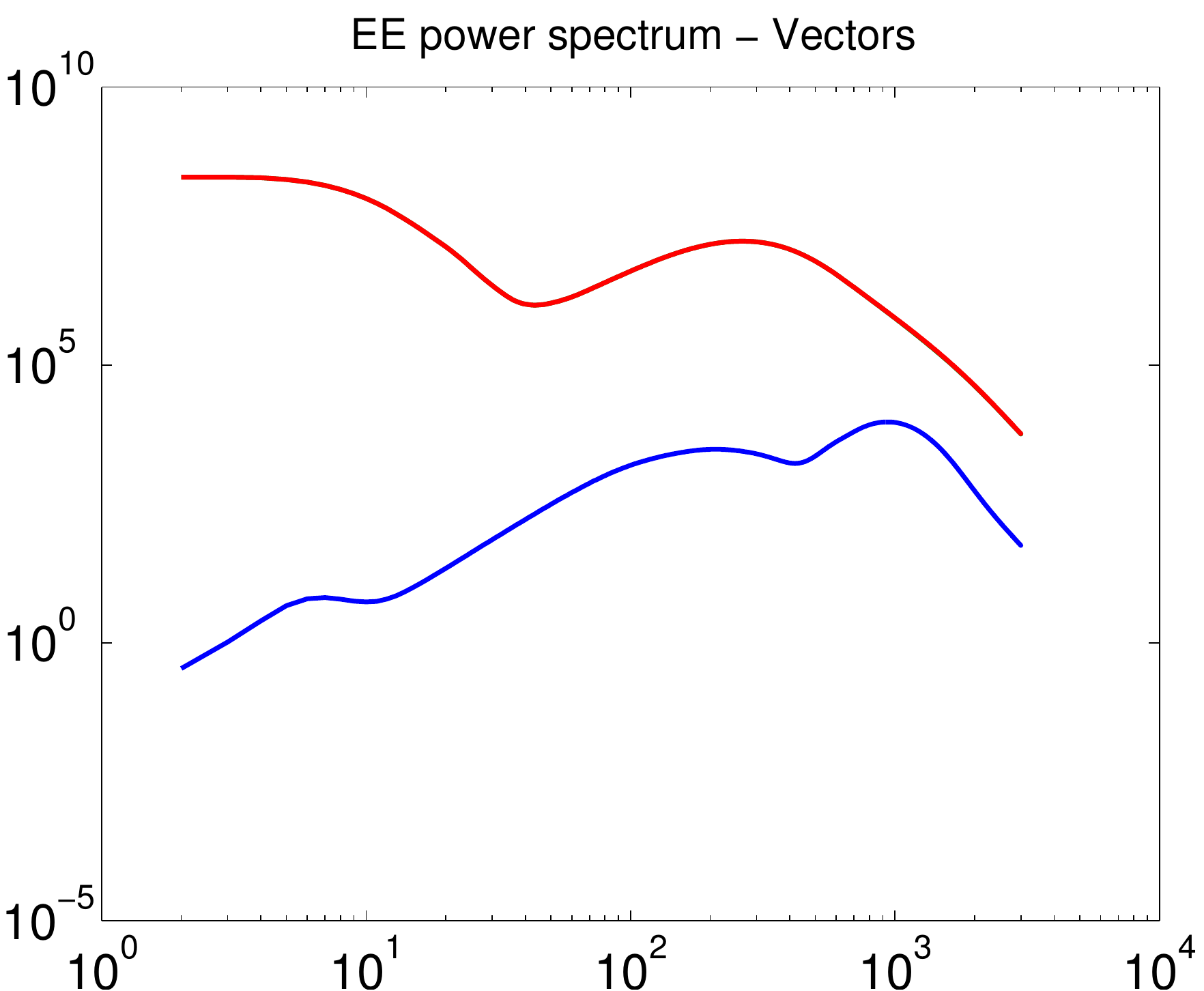} &
\includegraphics[width=1.63in]{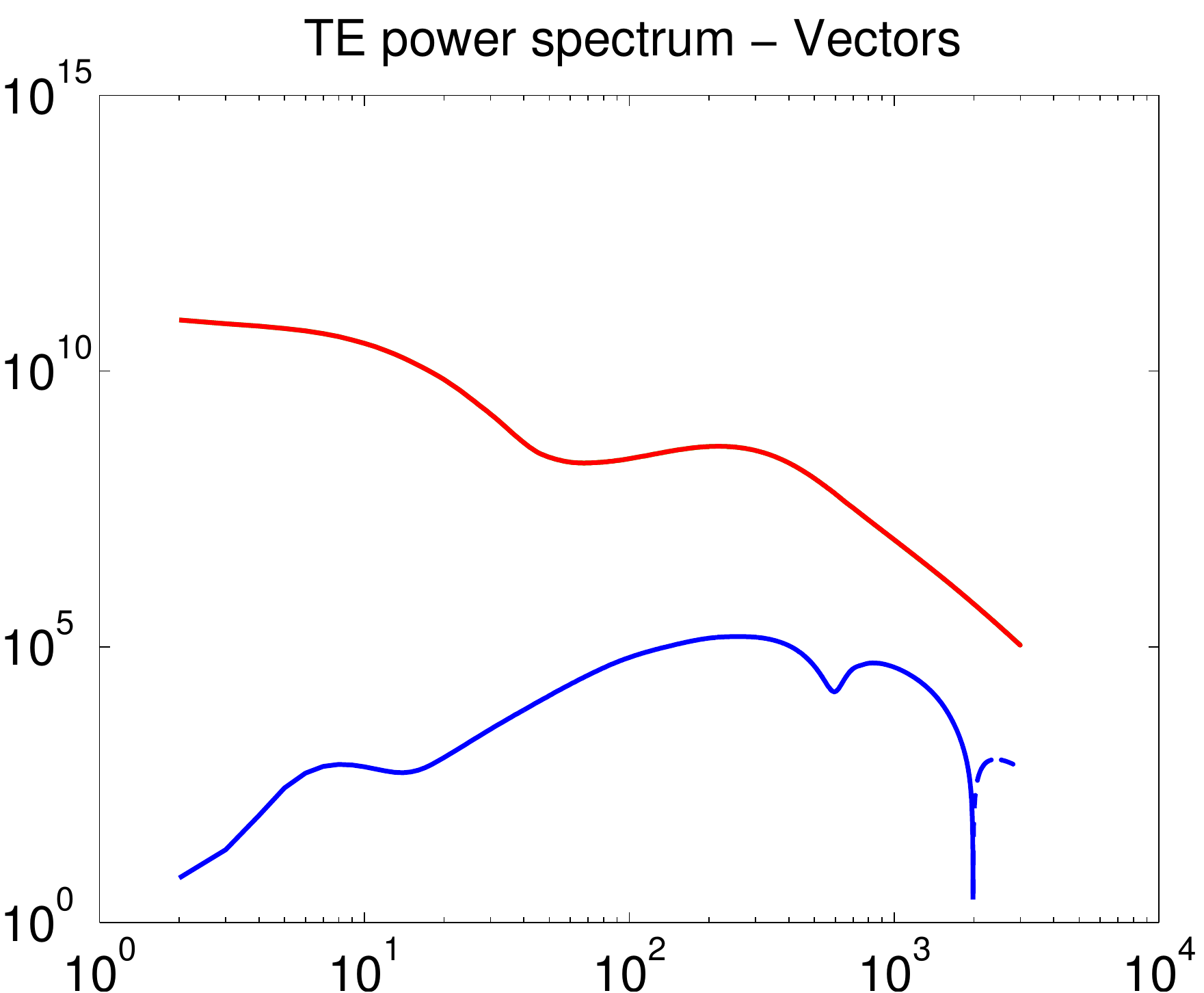} &
\includegraphics[width=1.63in]{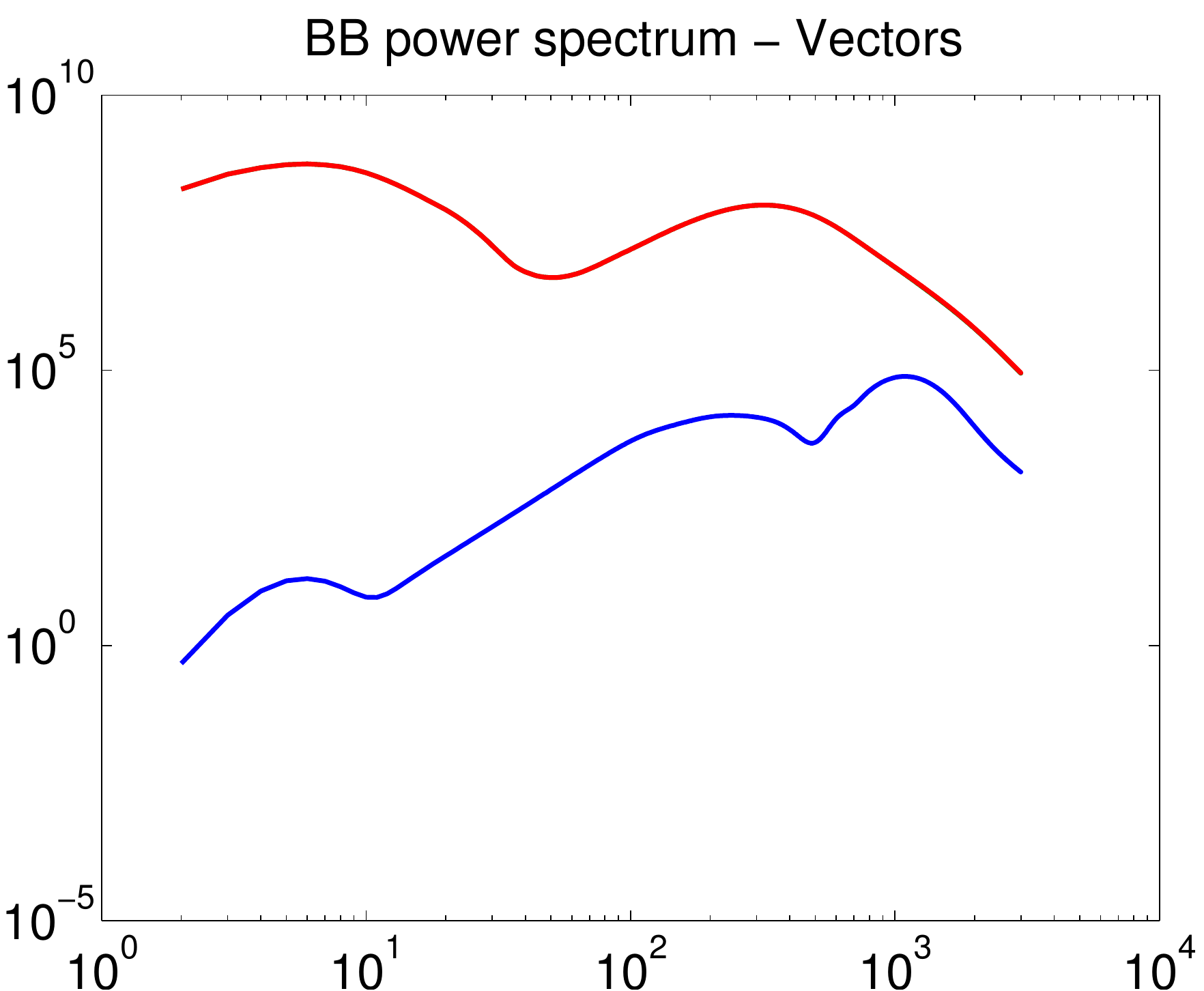} \\
\includegraphics[width=1.63in]{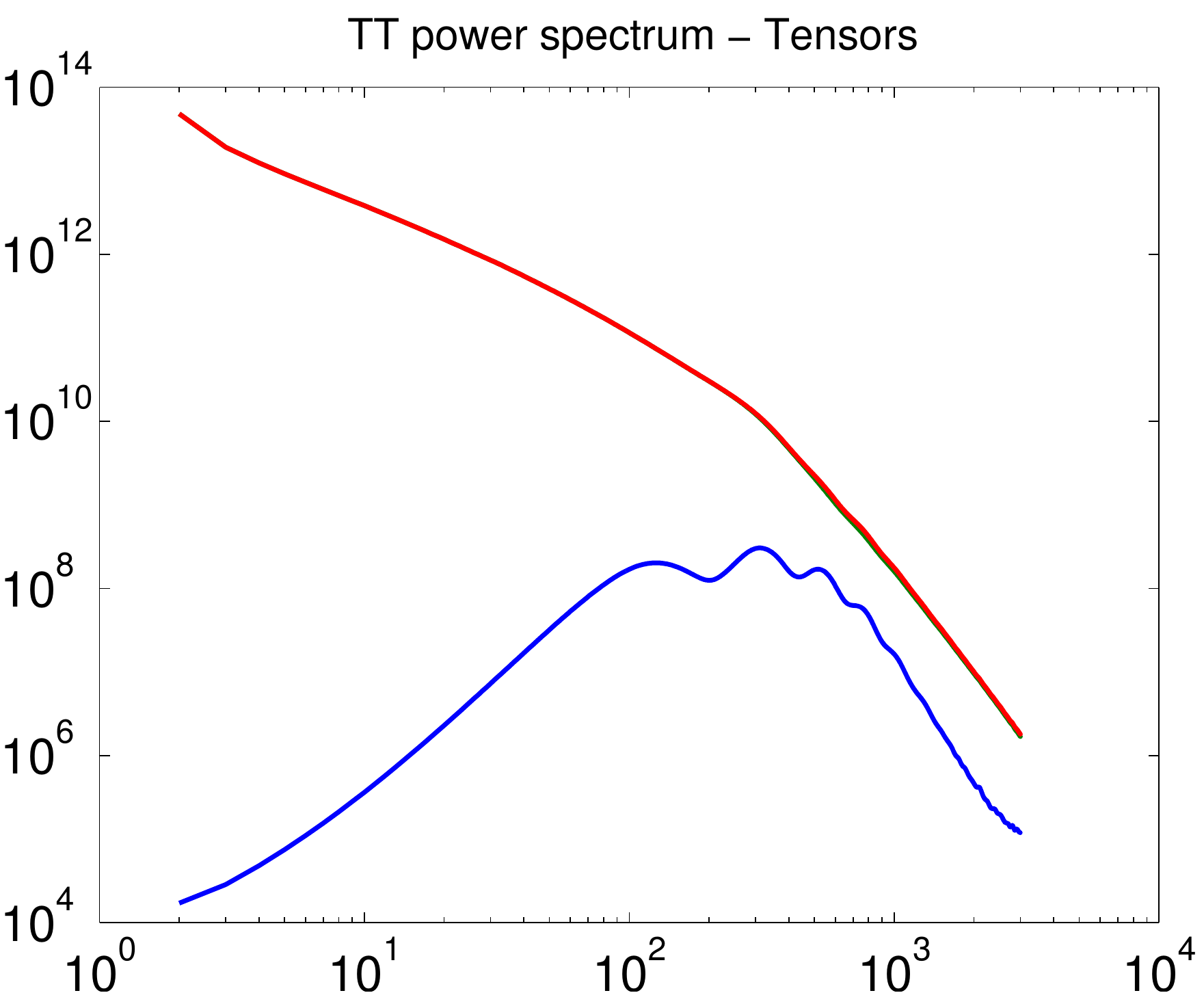} &
\includegraphics[width=1.63in]{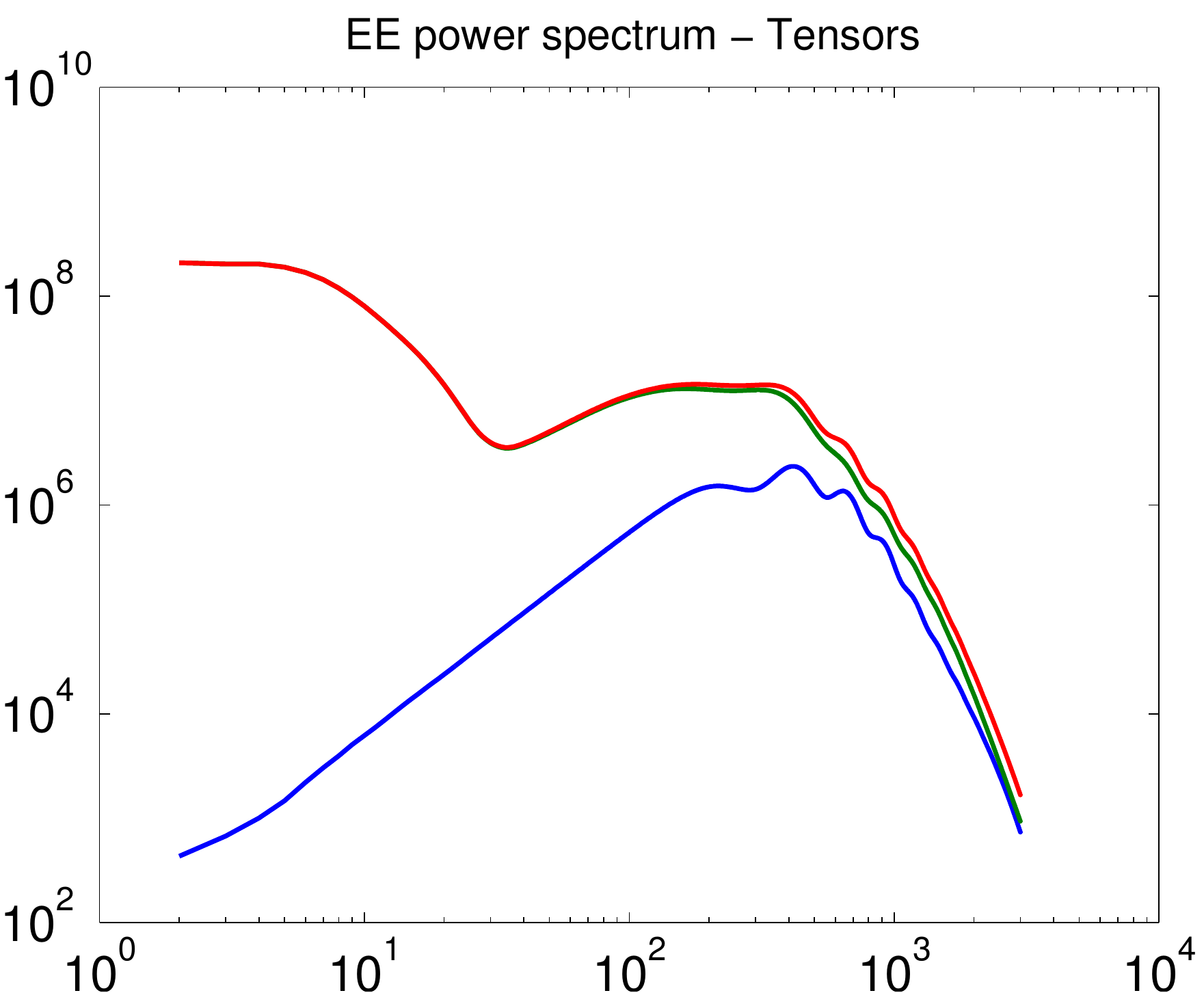} &
\includegraphics[width=1.63in]{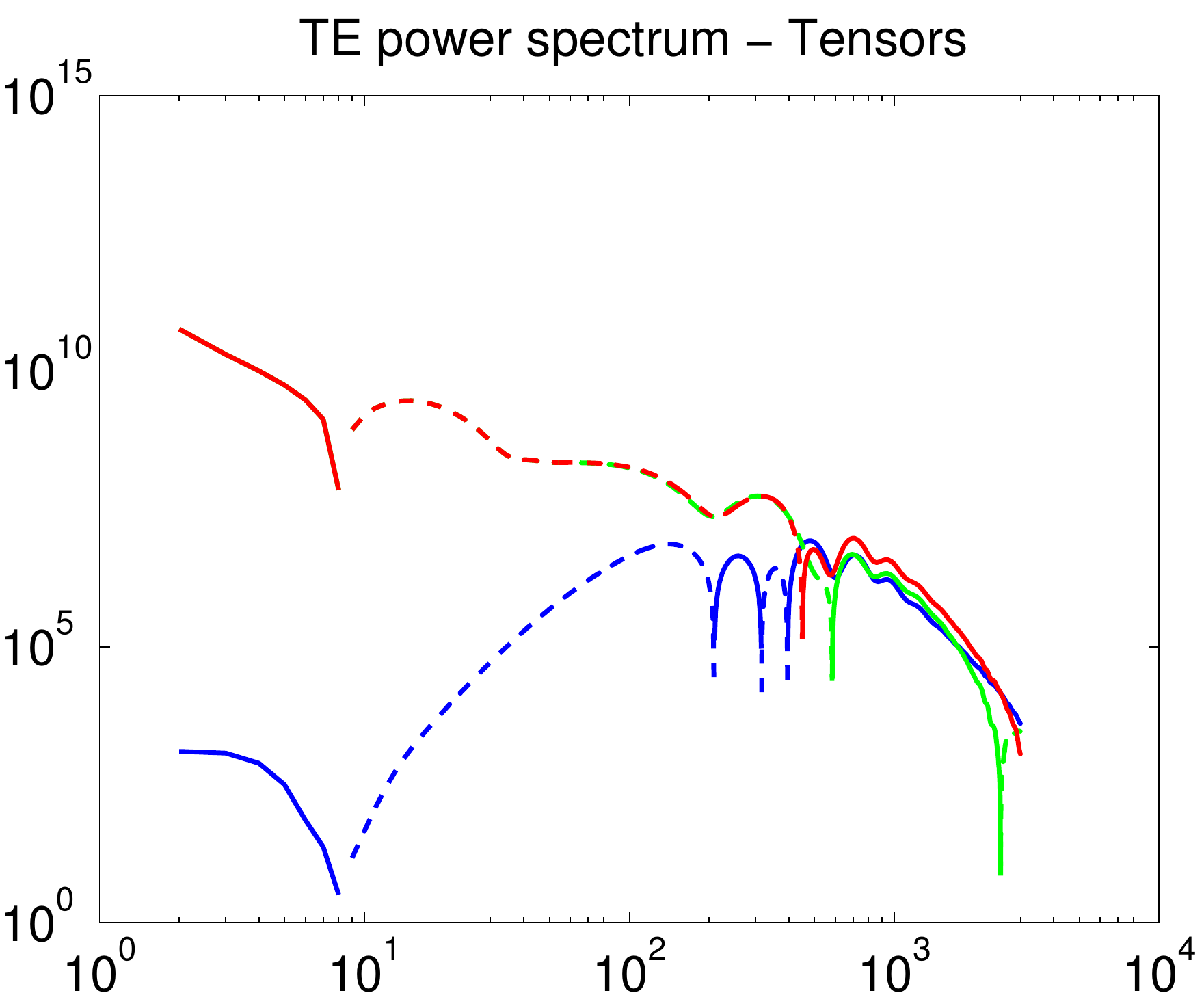} &
\includegraphics[width=1.63in]{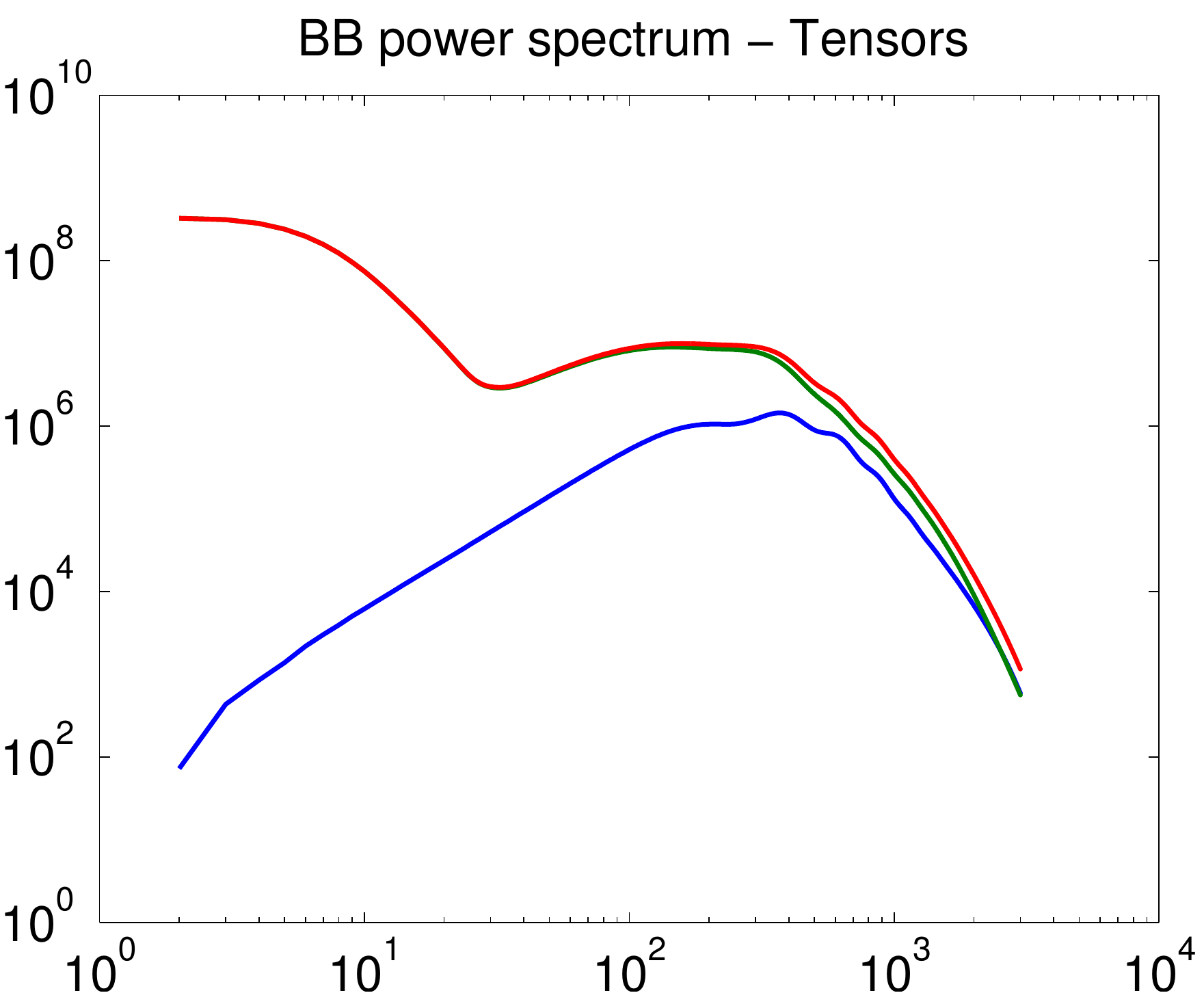} 
\end{array}$
\caption{Power spectra of domain walls showing the individual results from the radiation (in blue) and matter (in green) eras as well as total power spectra for scalars, vectors and tensors in the temperature and polarisation (EE, TE and BB) channels. In the case of the TE polarisation we have plotted the negative parts with dashed lines in the same colours as their positive counterparts. In the plots $\sigma/t_0=1.5 \times 10^{-7}$}
\label{mixed_ps_uetc}
\end{center}
\end{figure*}

We have calculated the power spectra for the domain walls in the temperature and polarisation channels for the radiation and matter era simulations separately and then by combining the two simulations together. We have also realised this in the three-simulations scenario.	

The results show that the radiation era contribution has a subdominant effect. This was expected, because the growth of the density of domain walls over time would mean that their most significant contribution is at late times. Indeed, as the matter era results completely dominate the power spectra, the errors from the procedure of combining the simulations become completely negligible. In Figure \ref{mixed_ps_uetc} we have plotted separately the contributions from the radiation and matter epochs to the power spectra in the temperature and polarisation channels showing how the matter era dominates on all the scales of interest. Only in the scalar TE plot one can see a more important effect of the radiation era. The fact that the power spectra are completely dominated by matter is in agreement to the domination of the results by the late-time contributions and also on the necessity to run a simulation with a higher expansion rate.

\begin{figure}[!htb]
\begin{center}$
\begin{array}{cc}
\includegraphics[width=1.5in]{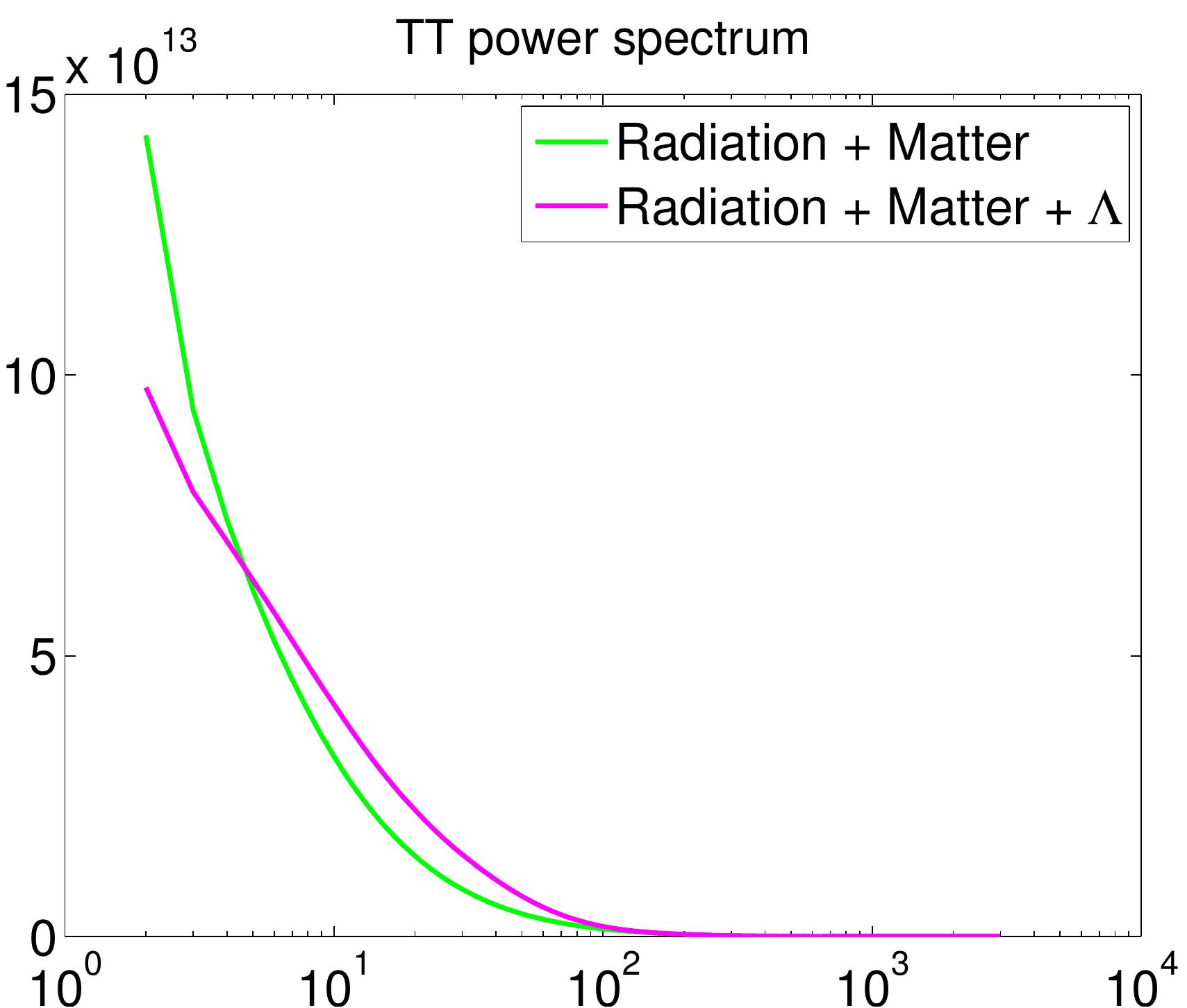} &
\includegraphics[width=1.5in]{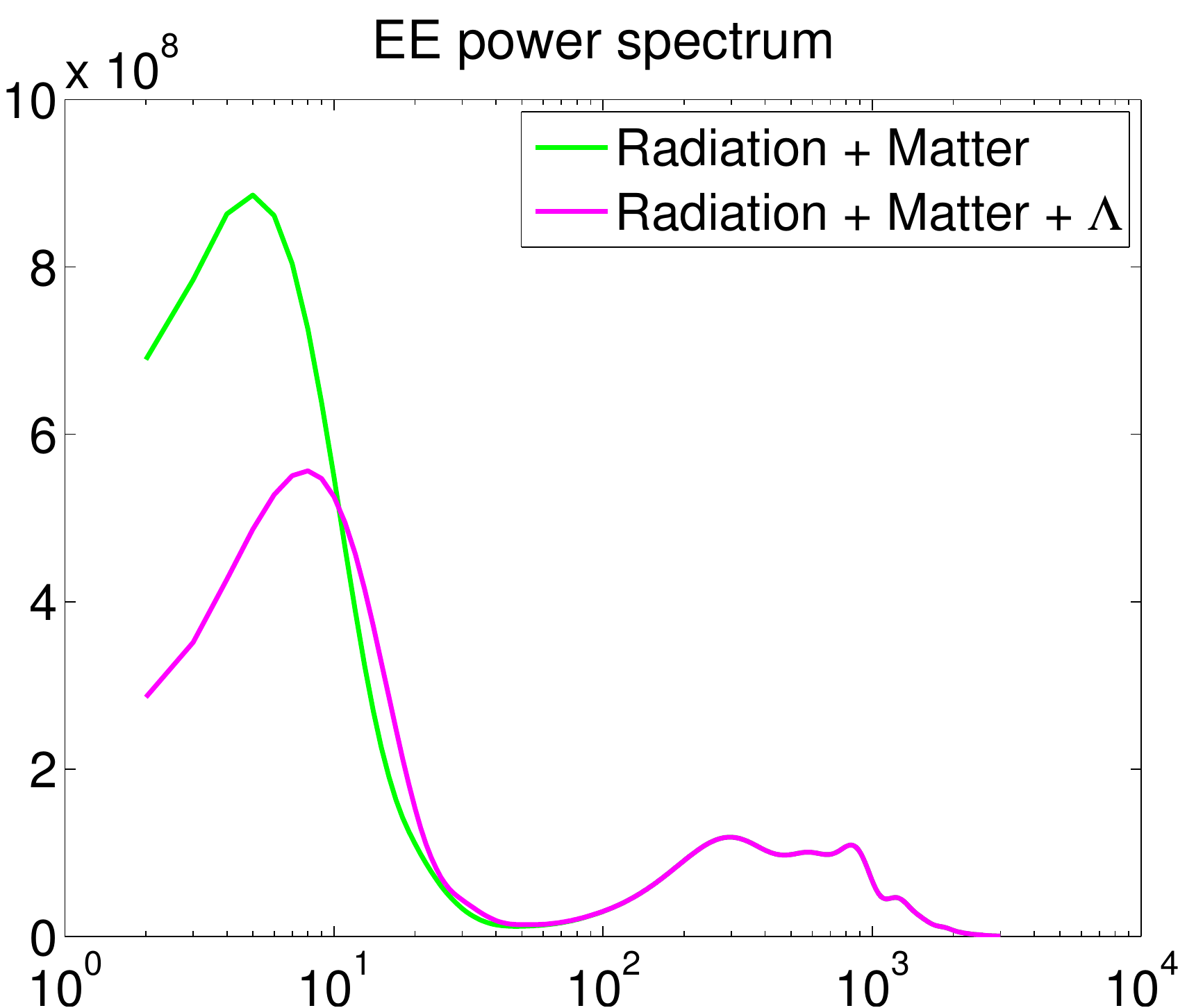} \\
\includegraphics[width=1.5in]{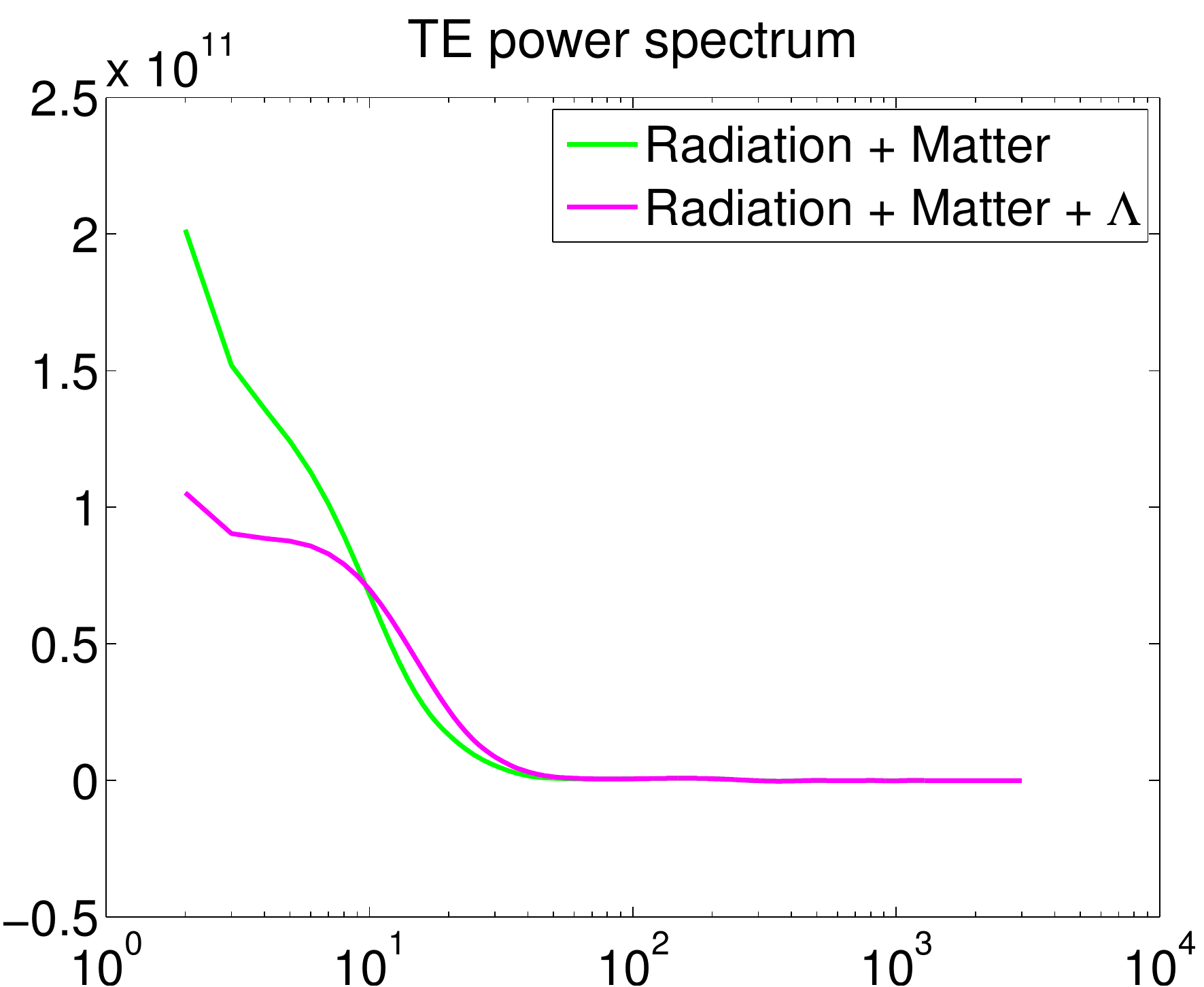} &
\includegraphics[width=1.5in]{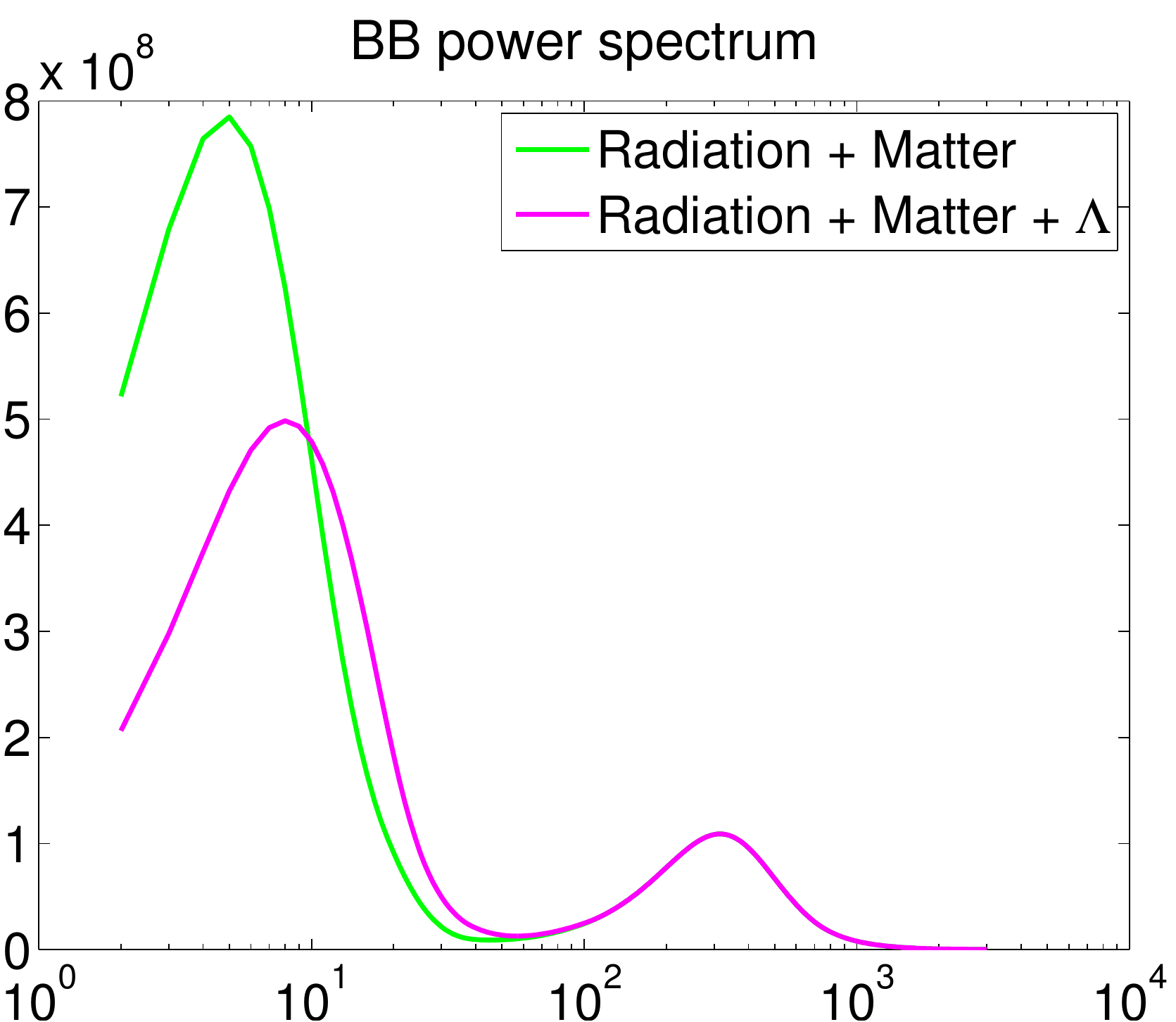} 
\end{array}$
\caption{Comparison between the domain walls power spectra from radiation \& matter eras and radiation, matter \& $\Lambda$ eras. In the plots $\sigma/t_0=1.5 \times 10^{-7}$}
\label{complambda}
\end{center}
\end{figure}

By adding the late-time simulation, the peak of the curves at $l=2$ drops by about one third in all four power spectra considered (Fig. \ref{complambda}). In that case, the expansion rate of the universe is faster than the growth of the domain wall density and hence the power is leaking to intermediate scales, keeping an approximately constant integrated power spectrum. As a consequence there is only a small change in the CMB constraint on domain walls in the two scenarios.

\section{CMB constraints on domain walls}
We used the COSMOMC code \cite{cosmomc2} which uses a Markov chain - Monte Carlo method to obtain constraints on the allowed contribution of the domain walls to the CMB power spectrum. We had to modify the code to accommodate the power spectrum from the domain walls. As the domain wall matter perturbations are uncorrelated to the primordial fluctuations, their power spectrum can be calculated separately. This is very helpful, because although their spectrum would depend on the cosmological parameters, the relative change to the inflationary spectrum would be small. For cosmic strings it has been checked \cite{kunz-cosmomc, battye-cosmomc} that this variation is less than 20\% and this is expected to happen for domain walls as well. Domain walls are tightly constrained by their TT power spectrum shape  and hence the parameter variation impact would not be very significant.
We have used the standard $\Lambda$CDM six-parameter model, together with a parameter quantifying the amplitude of the spectrum of the walls together with the latest Planck likelihoods.

We have analysed the radiation and matter scenario, and also one involving a late-time cosmological constant epoch. For the radiation and matter scenario, we have obtained a constraint on the surface density of the domain walls of $\sigma<4.22 \times 10^{-9}  \text{kg}/\text{m}^2$ (at 95\% confidence level), which corresponds to an energy scale of formation for domain walls of 0.96 MeV \cite{shbook}. 

By considering in addition the cosmological constant era, the constraints become $\sigma<3.85 \times 10^{-9}  \text{kg}/\text{m}^2$ and 0.93 MeV.
Both are in very good agreement with very rough observational constraints based just on the anisotropy constraint $\delta T/T \leq 10^{-5}$, which suggest that their energy scale should be less then 1 Mev (the original Zel'dovich bound) \cite{1975JETP...40....1Z}. 

Even though intuitively one may expect the constraint to weaken by adding the cosmological constant era (due to the fact that there is less power on very large scales), this does not happen because there is additional power on intermediate scales. There are large error for small \textit{l} and beyond $l=10$ the integrated power spectra are almost equivalent.

\begin{figure}[!htb]
\begin{center}
\includegraphics[width=2.7in]{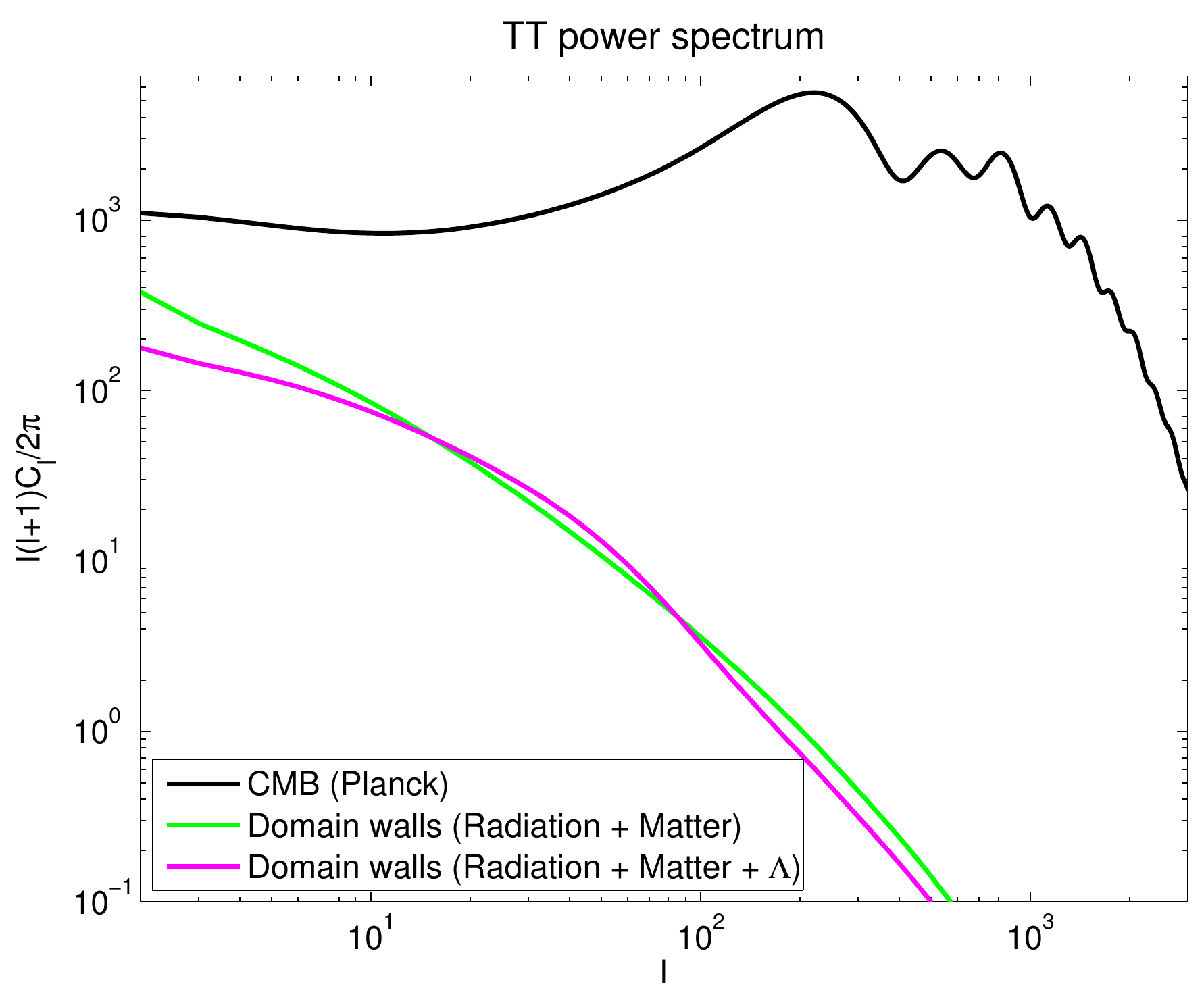} 
\caption{Comparison between the CMB power spectrum (from Planck) and the power spectrum from domain walls (normalised at the $2\sigma$ value of the energy scale in the two scenarios considered.}
\label{comp_planck}
\end{center}
\end{figure}

The values of the $\Lambda$CDM parameters do not shift significantly from the standard best fit $\Lambda$CDM Planck values, without domain walls. This is illustrated in Table \ref{table_params}. This is due to the fact that the allowed contribution of domain walls is very small, because of the different shape of their temperature power spectrum.

\begin{table}[!htb]
\small
\centering
\caption{Constraints on the fitted cosmological parameters, together with $1\sigma$ error bars in a full likelihood analysis (with all relevant nuisance parameters) with and without domain walls in the case of Planck and WMAP polarisation in the two domain walls scenarios considered.}
\begin{tabular}{|c|c|c|c|}
\hline
Parameter                           & No Walls           & Walls (R+M) & Walls (R+M+$\Lambda$)                 \\ \hline
\hline
$\sigma<$ (95\%)                    & -                  & 0.96  & 0.93       \\ \hline 
$H_0$                               & $67.20\pm 1.16$    & $67.25 \pm 1.18$    & $67.31 \pm 1.18$      \\ \hline 
$100\Omega_b h^2$                   & $2.202\pm 0.027$   & $2.201 \pm 0.028$   & $2.203 \pm 0.028$         \\ \hline
$\Omega_c h^2$                      & $0.120 \pm 0.003$  & $0.119 \pm 0.003$   & $0.119\pm 0.003$         \\ \hline
$\tau$                              & $0.089 \pm 0.013$  & $0.088 \pm 0.013$   & $0.088 \pm 0.013$    \\ \hline
$100\theta_{MC}$                    & $1.0412 \pm 0.0006$& $1.0412 \pm 0.0006$ & $1.0412 \pm 0.0006$        \\ \hline
$\ln(10^{10}A_s)$                   & $3.088 \pm 0.025$  & $3.085 \pm 0.025$  & $3.086 \pm 0.024$       \\ \hline
$n_s$                               & $0.959 \pm 0.007$  & $0.960 \pm 0.007$   & $0.960 \pm 0.007$            \\ \hline
\end{tabular}
\label{table_params}
\end{table}
Using these values of the energy scale, we have plotted on the same graph in logarithmic scale the standard CMB Planck power spectrum \cite{planckres} and the domain walls power spectra, normalised at the 95\% confidence level for its surface density (Fig. \ref{comp_planck}). The plot shows that indeed the domain walls only contribute on large scales as their power spectrum is quickly decaying in \textit{l}-space.

\section{Conclusions}
In this paper we have used high-resolution simulations based on the PRS algorithm to evaluate the energy-momentum tensor of a network of domain walls in an expanding universe, covering the radiation, matter and late-time $\Lambda$-domination eras. We have analysed how the wall network scales and we have then evaluated its unequal time correlator components in each epoch. We have used the rescaled eigenvectors and eigenvalues obtained from these correlators as sources into an Einstein-Boltzmann solver and we have thus determined the power spectrum of the domain wall network. The temperature power spectrum is quickly decreasing as a function of $l$ and has its maximum where the CMB measurements have large error bars. This allows the presence of some domain walls even though the shape of their power spectrum is completely different to the one of the CMB.

We have analysed two scenarios: one where only radiation and matter eras are considered and one which involves in addition a fast-expansion rate. We have shown that although there are noticeable changes in the power spectra that were obtained, the CMB constraints vary insignificantly.

We have used the CMB power spectrum to find the first precise quantitative constraint on the domain wall surface density, with an energy scale of 0.93 MeV at the 95\% CL for the standard $\Lambda$-cosmology.

\section{Acknowledgements}
AL is supported by STFC. CJM is supported by an FCT Research Professorship, contract reference IF/00064/2012, funded by FCT/MCTES (Portugal) and POPH/FSE (EC), and acknowledges additional support from the FCT grant PTDC/FIS/111725/2009. This work was undertaken on the COSMOS Shared Memory system at DAMTP, University of Cambridge operated on behalf of the STFC DiRAC HPC Facility. This equipment is funded by BIS National E-infrastructure capital grant ST/J005673/1 and STFC grants ST/H008586/1, ST/K00333X/1.

\bibliographystyle{model1a-num-names}
\bibliography{Bibliografie}{}

\end{document}